\documentclass[aps,prl,twocolumn,superscriptaddress,floatfix,citeautoscript,hyperref=pdftex]{revtex4-1}
\usepackage{graphicx}
\usepackage{booktabs}
\usepackage{dcolumn}
\usepackage{bm}
\usepackage{color}
\usepackage{amsmath}
\usepackage{amssymb}
\usepackage{multirow}
\usepackage{makecell}
\newcommand {\beq} {\begin{equation}}
\newcommand {\eeq} {\end{equation}}
\newcommand {\bqa} {\begin{eqnarray}}
\newcommand {\eqa} {\end{eqnarray}}

\usepackage[colorlinks=true]{hyperref}

\newcommand{\bk}{\bm{k}}

\begin{document}
\let\origaddcontentsline\addcontentsline
\renewcommand{\addcontentsline}[3]{}

\title{Superfluid weight in a Bardeen-Cooper-Schrieffer (BCS) model of  heavy and light fermions of twisted bilayer
graphene}

\author{Anushree Datta}
\email{anushree.datta@aalto.fi}
\affiliation{Department of Applied Physics$,$ Aalto University School of Science$,$ FI-00076 Aalto$,$ Finland}

\author{Eeli O.~Lamponen}
\affiliation{Department of Applied Physics$,$ Aalto University School of Science$,$ FI-00076 Aalto$,$ Finland}

\author{Haoyu Hu}
\affiliation{Department of Physics$,$ Princeton University$,$ Princeton$,$ NJ 08544$,$ USA}
\affiliation{Donostia International Physics Center (DIPC)$,$ Paseo Manuel de Lardiz\'{a}bal.~20018$,$ San Sebasti\'{a}n$,$ Spain}
\affiliation{Department of Physics, University of Science and Technology of China, Hefei, Anhui 230026, China}

\author{Pertti J.~Hakonen}
\affiliation{Department of Applied Physics$,$ Aalto University School of Science$,$ FI-00076 Aalto$,$ Finland}

\author{Dmitri K.~Efetov}
\affiliation{Fakult\"at f\"ur Physik, Ludwig-Maximilians-Universit\"at, Schellingstrasse 4, 80799 M\"unchen, Germany}
\affiliation{Munich Center for Quantum Science and Technology (MCQST), M\"unchen, Germany}

\author{B. Andrei Bernevig}
\affiliation{Department of Physics$,$ Princeton University$,$ Princeton$,$ NJ 08544$,$ USA}
\affiliation{Donostia International Physics Center (DIPC)$,$ Paseo Manuel de Lardiz\'{a}bal.~20018$,$ San Sebasti\'{a}n$,$ Spain}
\affiliation{IKERBASQUE$,$ Basque Foundation for Science$,$ 48013 Bilbao$,$ Spain}

\author{P\"aivi T\"orm\"a} 
\email{paivi.torma@aalto.fi}
\affiliation{Department of Applied Physics$,$ Aalto University School of Science$,$ FI-00076 Aalto$,$ Finland}

%TC:ignore
\begin{abstract}
We study superconductivity in twisted bilayer graphene (TBG) in a BCS formalism using the topological heavy fermion model. We calculate the superfluid weight (stiffness) and its quantum geometric contribution by self-consistently solving the BCS system assuming an attraction on top of the renormalized TBG bands with local self-energy. While the pairing resides in the heavy electrons of TBG, the light ones are found to be crucial for the stiffness. This arises from the heavy and light electron hybridization and concentration of quantum geometry near the $\Gamma$-point. Our calculated superfluid weight and transition temperatures are consistent with current experiments. The temperature dependence of the superfluid weight helps distinguish pairing gap structures, irrespective of the (yet unknown) mechanism and origin of the pairing.

\end{abstract}

\maketitle
%TC:endignore

{\it{Introduction.}}---TBG near the magic angle provides a minimal and highly tunable setting in which superconductivity emerges from narrow moiré bands rather than from conventional metallic dispersive bands. The superconducting domes observed upon doping appear in proximity to correlated insulating states, suggesting that pairing must be understood together with the strongly correlated normal state (NS)~\cite{Li2010Andrei,Bistritzer2011,Cao2018SC,Cao2018Insulator,balents2020superconductivity,andrei2021marvels,kennes2021,Torma2022}. A useful low-energy description is provided by the topological heavy fermion (THF) model, in which localized, strongly interacting ($f$) degrees of freedom centered near the AA stacking regions of TBG coexist and hybridize with more dispersive ($c$) carriers required by the topology of the flat bands~\cite{Jiang2019ChargeOrder,haule2019,Song2022,Calugaru2023,Shi2022,Datta2023,Rai2023,Chou2023,Hu2023,Lau2025}. This framework naturally accounts for both local-moment-like correlation physics and mobile charge transport. At the same time, TBG raises a fundamental question for flat single-particle-band superconductivity: how can a nearly dispersionless bare band support a sizable phase stiffness? Quantum geometry offers a possible answer - either by providing interaction-induced effective bandwidth or through multiband superfluid weight, since it can control both.~\cite{Peotta2015,Torma2022,Xie2020,Julku2020,Hu2019,Xie2020}. Recent transport, tunneling, and kinetic-inductance measurements in twisted graphene systems have sharpened this question by finding unconventional gap structure and superfluid stiffness scales that appear difficult to reconcile with a simple single-band Fermi-liquid picture~\cite{Oh2021,Tian2023,Tanaka2025,Banerjee2025}. 

In this Letter, we investigate superconductivity in TBG using the THF model within an effective Fermi-liquid description built on the correlated NS, and with assumed attractive interactions. We consider both onsite and nearest-neighbor (NN) pairing interactions and identify the possible superconducting symmetries. In particular, we find that NN pairing interactions yield several 
%of nearly degenerate 
superconducting states that are close in energy. By calculating the superfluid weight at zero and finite temperatures, we find that quantum geometry originating from the simultaneous presence of the heavy $f$ and light $c$ electrons is essential in describing the superfluid weight. While the starting model is derived from microscopic considerations, the strength of an added attractive interaction responsible for superconductivity is a free parameter, which we fix by matching the obtained gap value to experimental values. The calculated doping-dependent superfluid weight and the corresponding bare-stiffness Berezinskii-Kosterlitz-Thouless (BKT) transition temperature are within the range of experimentally measured  values~\cite{Tian2023,Tanaka2025,Banerjee2025}. There are many theories about the origin of SC in TBG \cite{Gao2026,Stepanov2020,Lau2025,Wang2024,Gonzalez2019,You2019, Wu2019, Lian2019, Khalaf2021, Shavit2021, Wang2021, Yu2023, Long2024,haule2019,Datta2023,Rai2023,Chou2023,Hu2023}; we are agnostic about its origin and instead investigate the \emph{consequences} of given pairing symmetries. 

\begin{table*}[ht]
\centering
\renewcommand{\arraystretch}{1.35}
\setlength{\tabcolsep}{7pt}
\begin{tabular}{|c|c|c|c|c|c|c|c|}
\hline
&
\makecell{Pairing\\interaction}
&
\makecell{$M^{\rm orb}\otimes M^{\rm val}$}
&
\makecell{Momentum\\dependence}
&
\makecell{Momentum\\parity (weight)}
&
\makecell{Symmetries\\preserved}
&
Irrep
&
\makecell{$\Delta \Omega$}
\\
\hline

SC1
&
\multirow{4}{*}{\makecell{Interorbital\\ NN}}
&
\makecell{
$\sigma_x\otimes\tau_x$\\
$i\sigma_y\otimes i\tau_y$\\
$\sigma_x\otimes i\tau_y$\\
$i\sigma_y\otimes\tau_x$
}
&
\makecell{
$d_{x^2-y^2}$\\
$d_{xy}$\\
$p_x$\\
$p_y+ if_{y(3x^2-y^2)}$
}
&
\makecell{Even ($\sim 12\%$)\\
Odd ($\sim 88\%$)}
&
$C_{2x},C_{2z}$
&
$A_1 \oplus E_2$
&
\makecell{0}
\\

\cline{1-1}\cline{3-8}

SC2
&
&
\makecell{
$\sigma_x\otimes\tau_x$\\
$i\sigma_y\otimes i\tau_y$\\
$\sigma_x\otimes i\tau_y$\\
$i\sigma_y\otimes\tau_x$
}
&
\makecell{
$d_{x^2-y^2}+id_{xy}$\\
$d_{x^2-y^2}-id_{xy}$\\
$p_x+ if_{y(3x^2-y^2)}$\\
$p_x- ip_{y}$
}
&
\makecell{Even ($\sim 8\%$)\\
Odd ($\sim 92\%$)}
&
$C_{2z}$
&
$A_1\oplus E_2$
&
\makecell{$0$}
\\

\cline{1-1}\cline{3-8}

SC3
&
&
\makecell{
$\sigma_x\otimes i\tau_y$\\
$i\sigma_y\otimes\tau_x$
}
&
\makecell{
$p_x$\\
$f_{y(3x^2-y^2)}$
}
&
Odd
&
$C_{2x},C_{2z}$, $\mathcal{T}$
&
$A_1 \oplus E_2$
&
\makecell{$\sim 0.02$ meV}
\\

\cline{1-1}\cline{3-8}

SC4
&
&
\makecell{
$\sigma_x\otimes\tau_x$\\
$i\sigma_y\otimes i\tau_y$
}
&
\makecell{
$d_{x^2-y^2}$\\
$d_{xy}$
}
&
Even
&
$C_{3z}$,$C_{2x},C_{2z}$,$\mathcal{T}$
&
$A_1$
&
\makecell{$\sim 0.05$ meV}
\\

\cline{1-1}\cline{3-8}

SC5
&
&
\makecell{
$\sigma_x\otimes i\tau_y$\\
$i\sigma_y\otimes\tau_x$
}
&
\makecell{
$p_x$\\
$p_y$
}
&
Odd
&
$C_{3z}$,$C_{2x},C_{2z}$,$\mathcal{T}$
&
$A_1$
&
\makecell{$\sim 0.05$ meV}
\\

\cline{1-1}\cline{3-8}

SC6
&
&
\makecell{
$\sigma_x\otimes\tau_x$\\
$i\sigma_y\otimes i\tau_y$\\
$\sigma_x\otimes i\tau_y$\\
$i\sigma_y\otimes\tau_x$
}
&
\makecell{
$d_{xy}$\\
$d_{x^2-y^2}$\\
$f_{y(3x^2-y^2)}$\\
$p_{x}$
}
&
\makecell{Even ($\sim 25\%$)\\
Odd ($\sim 75\%$)}
&
$C_{2x}$,$C_{2z}$,$\mathcal{T}$
&
$A_2\oplus E_2$
&
\makecell{$\sim 0.04$ meV}
\\

\hline

SC7
&
\multirow{3}{*}{\makecell{Intraorbital\\ NN}}
&
\multirow{2}{*}{\makecell{$\sigma_0\otimes\tau_x$}}
&
$d_{x^2-y^2}+i d_{xy}$
&
Even
&
$C_{2z},C_{3z}$
&
$E_2$
&
\makecell{$0$}
\\

\cline{1-1}\cline{4-8}

SC8
&
&
&
ext-$s$
&
Even
&
$C_{3z}$,$C_{2x},C_{2z}$,$\mathcal{T}$
&
$A_1$
&
\makecell{$\sim 0.18$ meV}
\\

\cline{1-1}\cline{3-8}

SC9
&
&
\makecell{
$\sigma_0\otimes i\tau_y$\\
}
&
$f_{y(3x^2-y^2)}$
&
Odd
&
$C_{3z}$,$C_{2x},C_{2z}$,$\mathcal{T}$
&
$A_2$
&
\makecell{$\sim 0.04$ meV}
\\

\hline
\end{tabular}
\caption{
Symmetry classification of the self-consistent SC
order parameters obtained for the different NN interactions at $\nu=-2.9$. The third column gives the combinations of $M_a^{\rm orb}$ and $M_b^{\rm val}$ present in the pairing matrix decomposition in terms of the Pauli matrices $\sigma_i$ and $\tau_i$ in orbital and valley space, respectively, for the spin-singlet pairing; the fourth
column gives the momentum-space lowest harmonics of the corresponding
$d_{a,b,c}(\mathbf{k})$ and the fifth column is the weight of even ($d$) and odd ($p$, $f$) parity components in the order parameter. 
The sixth and seventh columns list the
preserved symmetries and the irreps 
under the $D_6$ point group, respectively. The last column gives the difference in grand potential per moiré unit cell
$\Delta \Omega=\Omega-\Omega^{\rm min}$
relative to the lowest-grand-potential-solution obtained
for the same pairing interaction. The states SC1-SC3 and SC6 are nematic, i.e., break $C_{3z}$ symmetry. SC7 is chiral; $C_{2x}$ and $\mathcal{T}$ symmetries are broken.} %{\color{blue} factor of 2 check in energy }
\label{tab:pairing_symmetries}
\end{table*}

 {\it Model and method.}---
We describe TBG using the microscopic THF
model derived from the low-energy wave functions of the
Bistritzer-MacDonald continuum model~\cite{Song2022,Bistritzer2011} in the moiré lattice.
Its low-energy degrees of freedom consist, per spin and valley, of two
localized (heavy) $f$ orbitals centered at the AA regions and four
dispersive (light) $c$ orbitals. The $f$ orbitals carry approximately
$95\%$ of the flat-band spectral weight, while the $c$ electrons
hybridize with them near the moiré $\Gamma$-point, which is essential for the
nontrivial topology and quantum geometry of the flat bands. See End Matter for model details. 

The microscopic THF model interactions comprise onsite and NN $f$-$f$ repulsions, $c$-$c$ repulsions, and $f$-$c$ density and exchange interactions. We consider an effective Fermi-liquid NS and incorporate the dominant onsite $f$-$f$ interaction through the self-energy $\Sigma_f(\omega)\simeq\Sigma_f(0)+\omega(1-Z^{-1})$, where $Z$ is the quasiparticle renormalization. The annihilation operators for $f$-electrons with ($\tilde f$) and without ($f$) renormalization are related as $f\simeq\sqrt{Z}\,\tilde f$. With this, the $f$-$c$ hybridization and the
$f$-sector of the Hamiltonian are renormalized by $\sqrt{Z}$ and $Z$, respectively, and the remaining off-site interactions are treated at the Hartree level.

Since the microscopic origin of superconductivity in TBG remains unsettled and speculative~\cite{Gao2026,Stepanov2020,Lau2025,Wang2024,Isobe2018,Kennes2018,Gonzalez2019,You2019, Wu2019, Lian2019, Chichinadze2020,Khalaf2021, Shavit2021, Wang2021, Kozii2022, Chatterjee2022,Annica2022,Yu2023,Long2024,haule2019,Datta2023,Rai2023,Chou2023,Hu2023}, we introduce phenomenological attractive on-site and NN interactions between the coherent $f$ quasiparticles on the moiré lattice and treat them self-consistently within BCS theory. The resulting pairing term is
\begin{equation}
H_{\Delta}=
\frac{1}{2}\sum_{\mathbf{k},\alpha\beta,\eta\eta', ss'}
\Delta_{\alpha\beta}^{\eta\eta',ss'}(\mathbf{k})
\tilde f_{\mathbf{k}\alpha\eta s}
\tilde f_{-\mathbf{k}\beta\eta' s'}
+\mathrm{H.c.}
\end{equation}
Here, $\alpha, \beta =1, 2$ label the $f$ orbitals, $\eta, \eta'=\pm$ the valleys, and $s,s'$ the spins. We denote the $f$ orbital states in orbital $\alpha$ and valley $\eta$ as $|\alpha\eta\rangle$. The matrix
elements of the pairing order parameter (OP) $\Delta(\mathbf{k})$ are given by $\left[\Delta(\mathbf{k})\right]_{\alpha\eta s,\beta\eta' s'}
=
\Delta_{\alpha\beta}^{\eta\eta', ss'}(\mathbf{k})$, which is obtained self-consistently. For symmetry classification, we decompose $\Delta({\bf{k}}) =
\sum_{a,b,c} d_{a,b,c}({\bf{k}})
[M_a^{\rm orb}\otimes M_b^{\rm val}\otimes M_c^{\rm spin}]$, where $M_a^{\rm orb}$, $M_b^{\rm val}$ and $M_c^{\rm spin}$ are matrix bases in orbital, valley and spin subspaces, respectively, and $d_{a,b,c}(\bf{k})$ denotes the corresponding momentum dependence.

{\it{Pairing symmetries.}}-- We first investigate the symmetry properties of the self-consistent superconducting (SC) solutions obtained from our numerical calculations with NN interactions. The corresponding solutions for onsite interactions have been discussed in the literature \cite{Wang2024,Liu2024}, but we also provide them in the Supplementary Material (SM) (See Sec.~S5 of SM~\cite{SM}) for completeness. With NN interactions, we obtain several SC solutions which are all close in energy. In the following, we discuss only the SC states near the theoretical optimal filling ($\nu=-2.9$
~\footnote{Theoretical optimal doping is defined as the doping where $T_{\rm BKT}$ is the highest. Experimentally, the observed optimal filling is around $\nu\sim -2.5$. The experimental filling factor is controlled by the gate capacitance, and the filling remains fixed across the SC transition. Here $\nu$ labels the normal-state (NS) filling at the corresponding chemical potential, which has been kept fixed in our SC calculation.
%In our model, well-developed Fermi-liquid behavior occurs near $\nu\simeq-2.9$, where the quasiparticle weight is the largest in $-3 <\nu < -2$~\cite{Rai2023, Hu2026}. 
See Fig.~\ref{fig:figem1} in End Matter for comparison of NS and SC fillings and pairing order parameter as a function of SC fillings.}). %relative to neutrality per moiré unit cell). 
The SC gap functions are classified according to the irreducible representations (irrep) of the $D_6$ lattice point group, characterized by their properties under $C_{3z}$, $C_{2z}$, and $C_{2x}$ transformations~\cite{IMartin2018,Wu2019,Yu2023,Bernevig2025,Liu2024}.   

In all our calculations, we consider 
 intervalley pairing, which couples degenerate normal-state electrons at opposite momenta and constitutes the leading SC instability in random phase approximation~\cite{Hu2026}. Owing to a $U(2)$ spin-rotation
symmetry in each valley, intervalley spin-singlet and spin-triplet
pairing are symmetry-related and degenerate; we therefore focus on spin-singlet pairing. For corresponding spin-triplet pairing, the $C_{2z}$ eigenvalue is reversed and the irreps change as
$A_1\rightarrow B_1$, $A_2\rightarrow B_2$, and $E_2\rightarrow E_1$. 
In orbital space, pairing between the two $f$ orbitals can be intra- or interorbital, corresponding to diagonal or off-diagonal components of the gap matrix, respectively.

\begin{figure}
\centering
\includegraphics[width=0.48\textwidth]{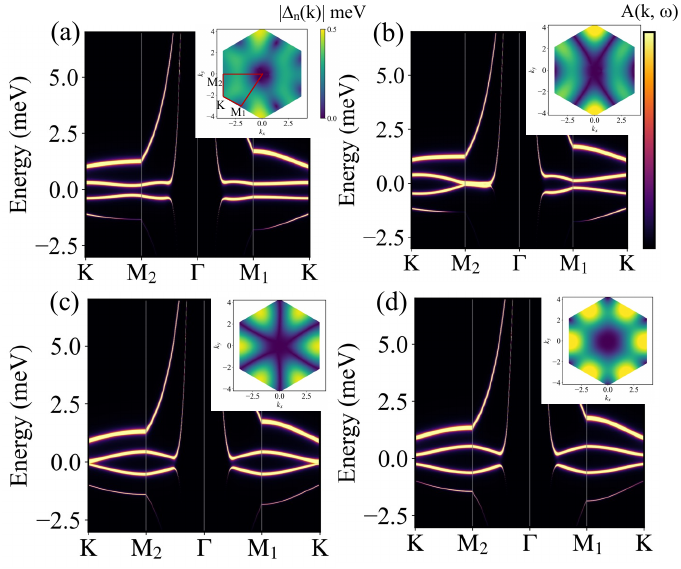}
\caption{(a-d) Electronic spectral functions $A(\mathbf{k},\omega)$ in the superconducting state for SC1 (a), SC3 (b), SC4 (c), and SC7 (d) states, along a high-symmetry path. Insets show the corresponding band-projected gap $|\Delta_n(\mathbf{k})|$ in the bottom flat band, $\Delta_n(\mathbf{k})=\sum_{\alpha\beta}u^{*}_{n\alpha+}(\mathbf{k})\Delta^{+-}_{\alpha\beta}(\mathbf{k})u_{n\beta+}(\mathbf{k})$. Here $u_{n\alpha\eta}(\mathbf{k})=\langle\alpha\eta|n_{\mathbf{k}\eta}\rangle$, with $|n_{\mathbf{k}\eta}\rangle$ denoting NS eigenstate $\alpha, \beta$ labeling $f$ orbitals.
} 
\label{fig:fig1}
\end{figure}

For interorbital NN pairing, we obtain several SC solutions with nearby grand potentials per moiré unit cell ($\Omega$), which we label SC1-SC6, and characterize in Table~\ref{tab:pairing_symmetries}. Among these SC1 and SC2 are nearly degenerate and have the lowest $\Omega$.
Both break $C_{3z}$ and time-reversal symmetry ($\mathcal{T}$), with gap functions containing a mixture of even- and odd-parity momentum harmonics and transforming as $A_1\oplus E_2$. Such parity mixing is consistent with fermionic antisymmetry, as the different momentum-parity components are accompanied by appropriate orbital-valley structures. The corresponding lowest-order harmonics, orbital-valley matrices, and relative weights of even and odd-momentum-parity components
are summarized in Table~\ref{tab:pairing_symmetries}. For the chosen interaction strength (see End matter), both SC1 and SC2 exhibit gap minima in the band-projected gap near the moiré $\mathrm{M}$ points, as shown in the band-projected gap and electronic spectral function, $A(\mathbf{k},\omega)$, in Fig.~\ref{fig:fig1}(a) for SC1. The gap minimum in the spectral function is more pronounced in SC2, as shown in Fig.~S4 of SM~\cite{SM}.

\begin{figure*}
\centering
\includegraphics[width=1.0\textwidth]{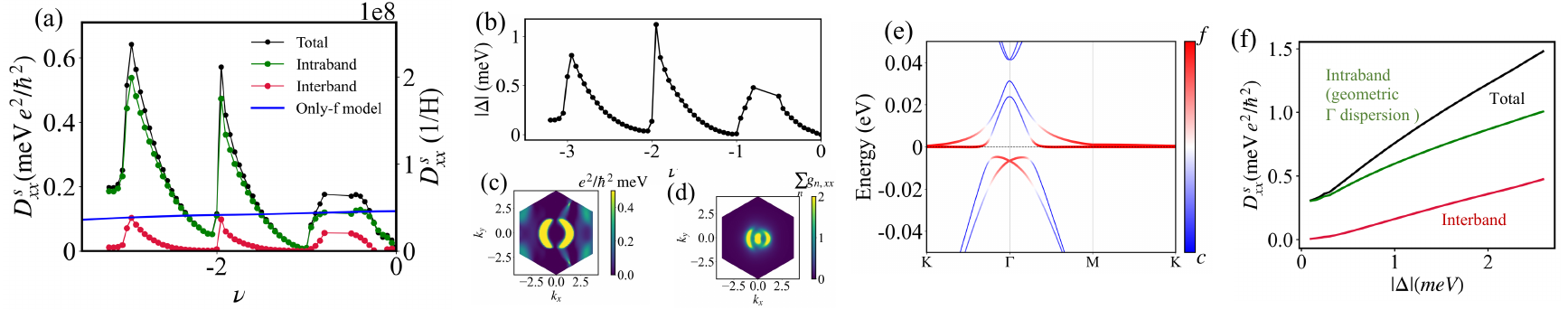}
\caption{(a) $D^s_{xx}$ and its inter- and intraband parts, and comparison of $D_s$ from a model with only $f$ electrons, as a function of the filling $\nu$. The left y-scale is in meV$e^2/\hbar^2$, and the right in 1/H (Henry). (b) Filling dependence of the magnitude of the pairing OP $|\Delta|$. (c) $D_{xx}^s$ integrand and (d) Quantum metric $\sum_{n\in\mathrm{flat}} g_{n,xx}(\mathbf{k})$ associated to coupling between correlated flat and remote bands, in the moiré BZ at $\nu = -2.9$. (e) Correlated NS band structure at $\nu = -2.9$, showing enhanced dispersion around the $\Gamma$-point. Colors indicate the $f$ and $c$ orbital weights. (f) $D^s_{xx}$ as a function of pairing OP, together with $D^{s, \rm intra}_{xx}$ and $D^{s, \rm inter}_{xx}$ at $\nu=-2.9$.}   
\label{fig:fig2}
\end{figure*}

The remaining solutions, SC3-SC6 (Table~\ref{tab:pairing_symmetries}), preserve $C_{2z}\mathcal{T}$ symmetry and have the grand potential $\Omega$ within $\sim 0.06$meV of SC1. Their band-projected gaps exhibit pronounced minima along $\mathrm M$ to $\Gamma$ direction in SC3, and along $\mathrm K$ to $\Gamma$ in SC4 [Inset, Figs.~\ref{fig:fig1}(b,c); see Figs. S7 and S8 in SM~\cite{SM} for SC5 and SC6]. These gap structures may be understood in terms of pairing nodes arising from the Euler obstruction for $C_{2z}\mathcal{T}$-symmetric intervalley pairing between same Chern number ($C$) sectors~\cite{Yu2023,Wang2024}, since the $f$-orbital basis approximately maps onto the Chern-band basis $C=(-1)^\alpha\eta$~\cite{Song2022,Liu2024}. These gap minima can yield low-energy Bogoliubov-de Gennes (BdG) quasiparticles depending on their position relative to the Fermi level and effects of interband pairing.

For intraorbital NN pairing, the lowest-$\Omega$ solution is a chiral $d$-wave state, denoted as SC7 ($E_2$ irrep) in Table~\ref{tab:pairing_symmetries}, where details of other competing solutions are also summarized. For onsite pairing, the lowest-$\Omega$, we obtain a nematic $E_2$ and $A_1$ states, for inter and intraorbital pairing, respectively~\cite{Liu2024,Wang2024} (see Sec.~S5 of SM~\cite{SM}).  
The distinct momentum-dependent gap anisotropies and nodal structures reflected in $A(\mathbf{k},\omega)$ (Fig.~\ref{fig:fig1}) for different pairing symmetries provide characteristic fingerprints that can be probed by momentum-resolved tunneling spectroscopy using the quantum twisting microscope~\cite{Waschitz2026}.

{\it{Superfluid weight and quantum geometry.}}--We now turn to the superfluid weight. We evaluate the superfluid-weight tensor $\mathbf D^s$, with components
$D_{\mu\nu}^{s}$ ($\mu,\nu$ are spatial indices), within the renormalized Fermi-liquid description
using the multiband BdG linear-response formalism
~\cite{Liang2017,lamponen2025super}.  In the normal-state 
eigenbasis $|n_{\bk \eta}\rangle$ of renormalized quasiparticle Hamiltonian $h^\eta_{\rm eff}$ (Eq.~\eqref{heff2} in the End Matter), with $\epsilon_{n\mathbf{k}\eta}$ denoting the corresponding quasiparticle energy, $n$ labels the
normal-state bands and $\eta=\pm$ the valleys, the current matrix elements are $[j^\eta_\mu(\mathbf{k})]_{nm}
=
\langle n_{\mathbf{k}\eta}|
\partial_{k_\mu}h^\eta_{\rm eff}
|m_{\mathbf{k}\eta}\rangle =
\delta_{nm}\partial_{k_\mu}\epsilon_{n\mathbf{k}\eta}
+
(\epsilon_{n\mathbf{k}\eta}-\epsilon_{m\mathbf{k}\eta})
\langle\partial_{k_\mu}n_{\mathbf{k}\eta}
|m_{\mathbf{k}\eta}\rangle$. In this basis, the superfluid weight can be written as

\begin{align}
    D^s_{\mu \nu } &=  \frac{2}{V}\sum_{\bk nmls} \bigg[  C_{++--}^{nmls} (\boldsymbol{k})[j_{\mu}^+(\boldsymbol{k})]_{nm}[j_{\nu}^-(-\boldsymbol{k})]_{sl} +(\mu\leftrightarrow\nu) \nonumber\\
    &+ \text{Re} \big\{ [\partial_\mu \Delta_{\bk}]_{nm} (C^{nmls}_{+-++}(\boldsymbol{k}) [j_{\nu}^{+}(\boldsymbol{k})]_{ls}
    \nonumber\\&+ C^{nmls}_{+---}(\boldsymbol{k}) [j_{\nu}^{-}(\boldsymbol{-k})]_{sl}) \big\} \bigg]
    % &- F^{nm}_{ls} ( [j_{\nu}^{+}(\boldsymbol{k})]_{ls} - [j_{\nu}^{-}(\boldsymbol{-k})]_{sl}),
    \label{eq: sfw_full}
\end{align}
where $[\partial_\mu \Delta_{\bk}]_{nm} = \langle n_{\bk +}| \partial_{\bk_\mu}\Delta_{\bk} | m^*_{-\bk -}\rangle$, $C_{p_1 p_2 p_3 p_4}^{nmls}(\boldsymbol{k})=\sum_{ab}\frac{n_{F}(E_{\bk a})-n_{F}(E_{\bk b})}{E_{\bk b}-E_{\bk a}}w_{p_1,an\bk}^{*}w_{p_2,bm\boldsymbol{k}}w_{p_3,bl\boldsymbol{k}}^{*}w_{p_4,as\boldsymbol{k}}$ with $p_i \in \{ +, -\}$, and $E_{\bk a}$ are the eigenenergies
of the BdG eigenstates $|\psi_{\mathbf{k}a}\rangle$, decomposed as $|\psi_{\mathbf{k}a}\rangle
=
\sum_{n=1}^{N_{\rm orb}}
\left[
w_{+,an\mathbf{k}}\,|+\rangle\otimes|n_{\mathbf{k}+}\rangle
+
w_{-,an\mathbf{k}}\,|-\rangle\otimes|n^{*}_{-\mathbf{k}-}\rangle
\right]$, with $|\pm\rangle$ the eigenvectors of the Pauli matrix $\gamma^z$ with eigenvalues $\pm 1$ in particle-hole space, $N_\text{orb}=6$, the number of orbitals per spin and valley, $V$ is the volume of the system, and $n_F$ is the Fermi-Dirac distribution. We decompose the superfluid weight as

\begin{equation}
D_{\mu\nu}^{s}
=
D_{\mu\nu}^{s,\mathrm{intra}}
+
D_{\mu\nu}^{s,\mathrm{inter}}. 
\end{equation}
The intraband contribution $D_{\mu\nu}^{s,\mathrm{intra}}$ is defined as the part of the sum in Eq. \eqref{eq: sfw_full} with $m=n$ and $l=s$, and thus contains current vertices diagonal in the normal-state basis, yielding the kinetic contribution associated with the band dispersion, as well as additional terms related to the $\bk$-dependence of $\Delta(\bk)$ that arise with NN interactions. The remaining terms define the interband contribution $D_{\mu\nu}^{s,\mathrm{inter}}$. With onsite interactions and in the isolated flat-band limit, $D_{\mu\nu}^{s,\mathrm{inter}}$ becomes proportional to the Brillouin-zone-integrated minimal quantum metric~\cite{Peotta2015,huhtinenRevisitingFlatBand2022}. 

Fig.~\ref{fig:fig2}(a) shows the 
longitudinal component of superfluid weight $D^s_{xx}$ ($D^s_{xx}\approx D_s =\sqrt{\det\mathbf D^s}$ for SC1, with $D^s_{yy}$ differing from $D^s_{xx}$ by $\approx 5\%$; see Fig.~S20 of SM~\cite{SM}) and its intra- and interband contributions, as functions of filling for SC1 (Table~\ref{tab:pairing_symmetries}). We choose SC1 as representative since it is the lowest-energy interorbital NN solution and remains stable throughout the filling range considered. Other SC states stable across this range exhibit qualitatively similar filling dependence (Fig.~S16 in SM~\cite{SM}).  

$D^s_{xx}$ exhibits a dome-like filling dependence between consecutive integer fillings, a feature consistent with experiments~\cite{Tian2023,Tanaka2025,Banerjee2025}. The maximum value of the superfluid weight occurs near $\nu \simeq- 2.9$, and its value is $2.4 \times 10^{8}$ 1/H. This is in good agreement with the maximal superfluid weight of
%$0.5 \times 10^{8}$ 1/H~\cite{Tian2023} 
$\sim1.7 \times 10^{8}$ 1/H from kinetic inductance
measurements in TBG~\cite{Tanaka2025}.
We note that the SC gap value for TBG is not precisely known; therefore, we use the value $0.4$ meV as an estimate of coherence-peak energy in the SC gap ~\cite{Nadjperge2026, pablo2026, Oh2021}, and adjust pairing interactions such that the corresponding minimum-energy states reproduce this gap at $\nu = -2.9$. 

$D^s_{xx}$ minima follow the nonmonotonic filling dependence of the pairing OP, $|\Delta|=\sum_{\ell,\alpha\beta}
|\Delta_{\alpha\beta}^{(\ell)}|$ where $\ell$ labels the NN bonds (Fig.~\ref{fig:fig2}(b)), as well as the quasiparticle weight $Z$ associated with $f$ orbitals in the Fermi liquid NS (See Fig.~S1 in SM~\cite{SM})
%inset of Fig.~\ref{fig:fig2}(b)). 
However, the maximum values of $D^s_{xx}$ and $|\Delta|$ occur at different fillings: $|\Delta|$ is largest for $\nu=-1.9$ but $D^s_{xx}$ is largest for $\nu=-2.9$.

Decomposition of $D^s_{xx}$ into inter- and intraband components reveals that the superfluid weight is dominated by the intraband contribution. $D^{s,\rm intra}_{xx}$ accounts for $\sim68\%$ of $D^s_{xx}$ near $\nu\simeq-0.8$ and $\sim83\%$ near $\nu\simeq-2.9$ [Fig.~\ref{fig:fig2}(a)]. The dominant $D^{s,\rm intra}_{xx}$ mainly originates from the Hartree-induced flat-band (narrow) dispersion upon doping [Fig.~\ref{fig:fig2}(e)], due to the presence of $c$ electrons near moiré $\Gamma$. The $c$ electrons are also essential for the nontrivial quantum geometry of the flat bands. Thus, $D^{s,\rm intra}_{xx}$ is intrinsically connected to the quantum geometry of the flat bands.

To establish this connection analytically, we neglect terms involving $\partial_{k_\mu}\Delta(\mathbf{k})$ ($\approx 5\%$ of the total in SC1) and the contribution from interband pairing in Eq.~\eqref{eq: sfw_full}, which are numerically found to be small, and write the intraband contribution as $D_{\mu\mu}^{\rm{s, intra}}
\simeq
\sum_{\mathbf{k},n}
\frac{
|\Delta_n(\bf{k})|^2
}{
\left[
\epsilon^2_{n\mathbf{k}+}
+
|\Delta_n(\mathbf{k})|^2
\right]^{3/2}
}
\,
\mathcal{V}_{n, \mu\mu}(\bf{k})$, where $\Delta_n(\bf{k})$ is the band-projected gap, and $\mathcal{V}_{n\mu\mu}(\mathbf{k})=([j^+_{\mu}(\mathbf{k})]_{nn})^2$. The connection to quantum geometry then takes 
a particularly simple form by further neglecting the bare flat-band bandwidth ($\simeq3.7\mathrm{meV}$) and Hartree-induced mixing between flat and remote bands. 
Projecting the Hartree potential onto these ideal bare flat-band subspace of the THF model, the dispersion of the flat bands (narrow) becomes $\epsilon_{n\mathbf{k}+}\simeq\epsilon_c +(\epsilon_f-\epsilon_c) w_{nf}(\mathbf{k})$, where \(\epsilon_f\) and \(\epsilon_c\) are the Hartree shifts associated with the \(f\) and \(c\) orbitals, respectively, and $w_{nf} (\bk)=\sum_{\alpha=1,2}|u_{n\alpha+}(\bk)|^2$ is the $f$-orbital weight in correlated flat band $n$. This gives 
$\mathcal V_{n,\mu\mu}(\mathbf{k})
\simeq
(\epsilon_f-\epsilon_c)^2
\left(\partial_{k_\mu} w_{nf}(\mathbf{k})\right)^2$.
By the Cauchy-Schwarz inequality, the $k$-derivative of $w_{nf}(\mathbf{k})$ satisfies
$\left(\partial_{k_\mu} w_{nf}(\mathbf{k})\right)^2
\leq
4 w_{nf}(\mathbf{k})w_{nc}(\mathbf{k})
g_{n,\mu\mu}(\mathbf{k})$,
where $g_{n,\mu\mu}$ is the quantum metric associated with transitions from the correlated flat-band $n$ to the remote-band subspace of the THF Hamiltonian (Details given in Sec.~S6 of SM~\cite{SM}) and $w_{nc}(\mathbf{k})= 1-w_{nf}(\mathbf{k})$ is the $c$ electron weight. Thus, the velocity tensor is bounded by the quantum metric,
$\mathcal{V}_{n, \mu\mu}(\mathbf{k})
\lesssim
4(\epsilon_f-\epsilon_c)^2
w_{nf}(\mathbf{k})w_{nc}(\mathbf{k})
g_{n,\mu\mu}(\mathbf{k})
$.
Hence, the dominant contribution to $D^{s,intra}_{\mu\mu}$ from band $n$ requires both a finite $g_{n,\mu\mu}$ and coexisting $f$- and $c$-orbital weights. Indeed, at $\nu= -2.9$, integrand of $D^s_{xx}$ in the moiré Brillouin zone (BZ) peaks in the vicinity of the $\Gamma$ point, where the quantum metric is concentrated and the flat bands exhibit mixed $f$ and $c$ character [Figs.~\ref{fig:fig2}(c)-(e)].

The role of quantum geometry in intraband superfluid weight can be further illustrated numerically by a reduced model with only the two $f$-orbitals per spin and valley, with parameters chosen to mimic the low-energy $f$-electron bandwidth of the full model~(also see End matter). 
In this reduced model, the Hartree potential only shifts the band energies by a constant and does not modify the bandwidth or eigenvectors.
As a result, even with a finite dispersion introduced in the $f$ sector, the superfluid weight remains much smaller than the maximal values in the full model with $f$ and $c$ electron hybridization, as shown in Fig. \ref{fig:fig2}(a).

Turning to the interband contribution, which also encodes quantum geometry through the off-diagonal current matrix elements $[j^\eta_\mu(\mathbf{k})]_{nm}$ with $n\neq m$, we find that $D^{s,\mathrm{inter}}_{xx}$ is dominated by mixed terms involving one diagonal and one off-diagonal current matrix element ($n \neq m$ and $l= s$, or $n = m$ and $l\neq s$ in Eq.~\eqref{eq: sfw_full}). The corresponding coherence factors $C_{p_1 p_2 p_3 p_4}^{nmls}$ vanish for purely intraband pairing, revealing the role of interband pairing in this contribution in TBG.

We now show $D^{s}_{xx}$ as a function of pairing OP $|\Delta|$ at $\nu = - 2.9$ (Fig.~\ref{fig:fig2}(f)). In the isolated-flat-band regime, the quantum-geometric contribution $D^{s, \rm inter}_{xx}\propto |\Delta|$, for 
$|\Delta|\ll E_{\rm gap}$ (the flat and remote bands separation)~\cite{Peotta2015,julku2016geometric,Torma2022}. In contrast, in a conventional single-band case at weak coupling, the intraband response is essentially independent of the pairing strength. Here, we find a nearly linear scaling of $D^s_{xx}$ with $|\Delta|$, dominated by 
the intraband contribution, within the relevant range of pairing interaction.

\begin{figure}
\centering
    \includegraphics[width=0.5\textwidth]{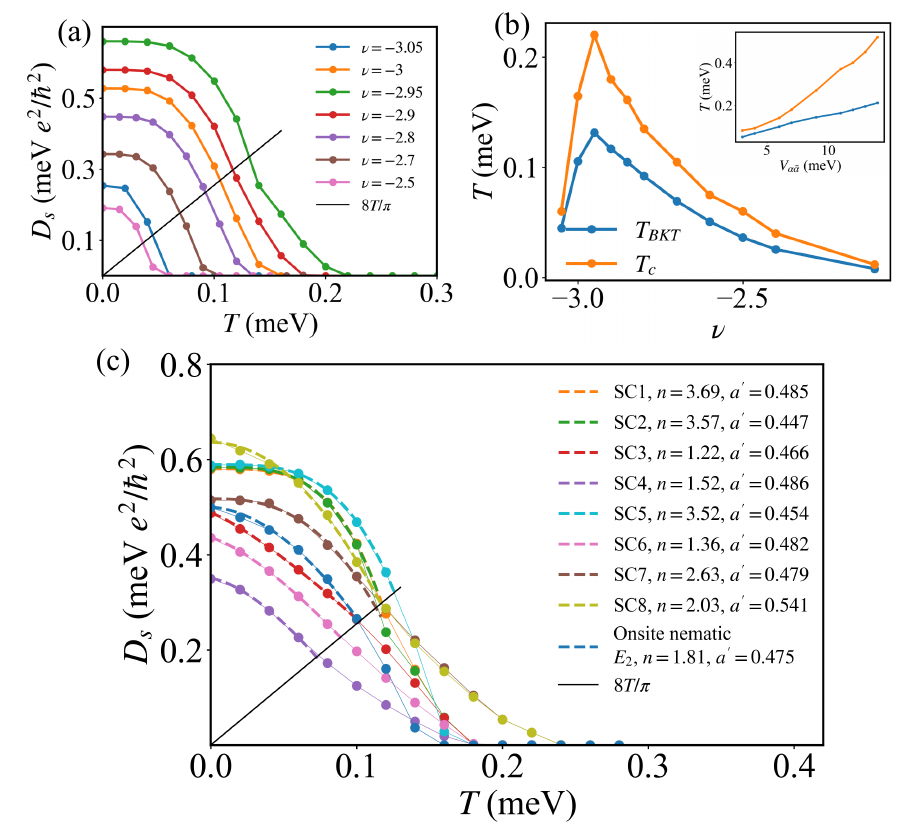}
\caption{(a) Temperature dependence of $D_s$ at different fillings for SC1, and the line (black) defining $T_{\mathrm{BKT}}$. (b) Corresponding $T_{\mathrm{BKT}}$ and $T_c$ as functions of filling. Inset presents $T_{\mathrm{BKT}}$ and $T_c$ at $\nu = -2.9$ with varying pairing interactions. (c) Temperature dependence of $D_s$ for different SC states and power-law fits below $T_{\mathrm{BKT}}$ using $D_s(T)=D_s(0)-aT^n$ at $\nu=-2.9$. Here $n$ is the fitted exponent and $a'=aT_{\mathrm{BKT}}^n/D_s(0)$.}
\label{fig:fig3}
\end{figure}

\begin{figure}
\centering
\includegraphics[width=0.4\textwidth]{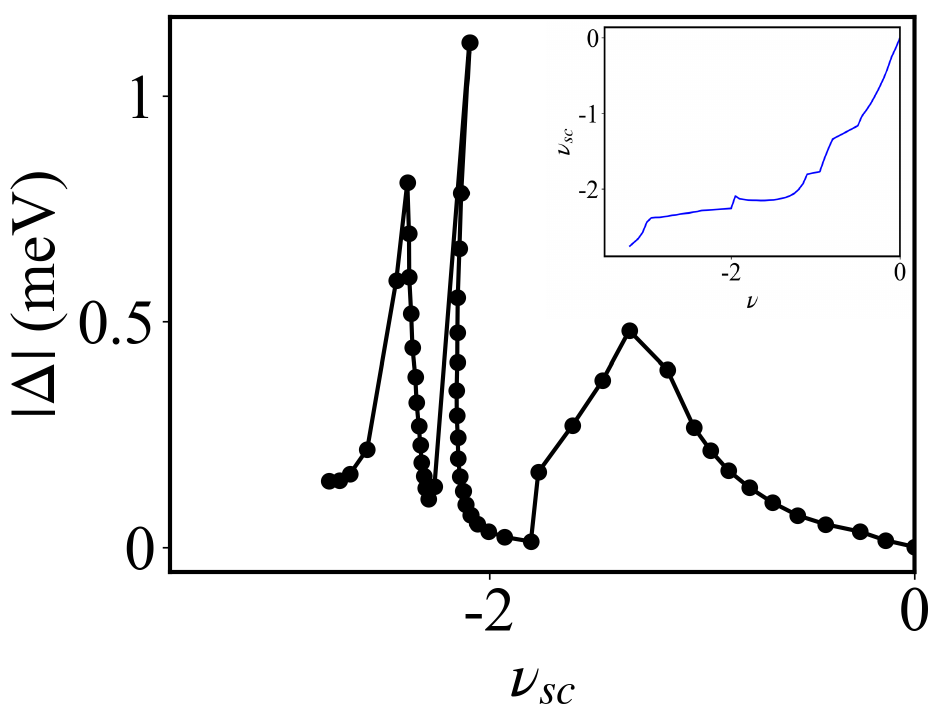}
\caption{Pairing order parameter as a function of $\nu_{SC}$. At each point, Hartree potentials and quasiparticle
weights are kept frozen at corresponding normal state values. Inset: SC state filling $\nu_{SC}$ as a function of NS filling factor $\nu$.}  
\label{fig:figem1}
\end{figure}

{\it{Finite temperature behavior and critical temperatures.}}--- 
We next examine the $D_s =\sqrt{\det\mathbf D^s}$ at finite temperatures. In Fig.~\ref{fig:fig3}(a), we show the evolution of $D_s$ as a function of temperature for doping values ranging from $\nu=-2$ to $-3$ for SC1. From the $D^s(T)$ curves, we estimate the SC transition temperature $T_{\rm c}$ from the vanishing of pairing OP, and the BKT transition temperature $T_{\mathrm{BKT}}$ using the BKT criterion: $T_{\mathrm{BKT}}=\frac{\pi}{8}D_s(T_{\mathrm{BKT}})$ for the bare stiffness without vortex renormalization, and neglecting possible higher than 2$e$ charge order \cite{Khalaf2022}. In Fig. \ref{fig:fig3}(b),
we show the variation of $T_{\rm BKT}$ and $T_{\rm c}$ as functions of doping, which exhibit a dome-like non-monotonic behavior, with the optimal $T_{\rm BKT}$ reaching approximately $1.5$ K. In Fig. \ref{fig:fig3}(b) inset, we vary the pairing interactions near $\nu=-2.9$ such that the SC gap ranges up to $\simeq 1$meV. Within this range, the $T_{\mathrm{BKT}}$ reaches up to $\simeq 2.3$ K, and $T_{\rm c}/T_{\rm BKT}$ grows with pairing interactions. By contrast, the only-$f$ model yields much smaller $T_{\rm BKT}\simeq 0.4$K (Fig.~S17 of SM~\cite{SM}). 

\begin{figure}
\centering
\includegraphics[width=0.4\textwidth]{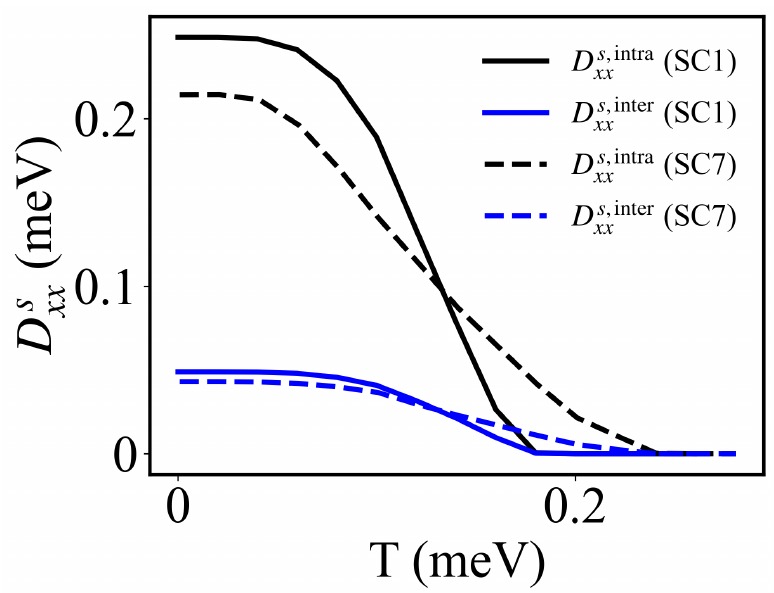}
\caption{Temperature dependence of $D^{s, \rm intra}_{xx}$ (black) and $D^{s, \rm inter}_{xx}$ (blue) contributions to $D^s_{xx}$ for SC1 (solid) and SC7 (dashed), lowest-$\Omega$ states for inter- and intraorbital pairing, respectively.}  
\label{fig:figem2}
\end{figure}

The temperature dependence of the superfluid weight also provides a useful probe of the pairing symmetry, since different pairing symmetries are associated with distinct quasiparticle spectra and hence different thermal suppression of $D_s$. In Fig.~\ref{fig:fig3}(c), we show the evolution of $D_s$ with temperature for different pairing states listed in Table~\ref{tab:pairing_symmetries} at $\nu = -2.9$, together with low-temperature power-law fits $D_s(0)-D_s(T)=a T^n
$ below $T_{\mathrm{BKT}}$, where $a$ is a fitting coefficient and $n$ is the power law exponent (see SM~\cite{SM} for fit-parameter analysis, RMSE and $R^2$, and corresponding results at $\nu = -1.9$). The states SC3, SC4, and SC6, characterized by pronounced gap minima in the band-projected gaps, exhibit smaller fitted exponents, with $n\simeq1.22$, $1.52$ and $n\simeq1.36$, respectively. In contrast, the nodeless SC1 and SC7 states show weaker thermal depletion, with $n\simeq3.69$ and $2.63$, respectively. SC2 and SC5, where gap minima are located in regions of suppressed $D^s_{\mu\mu}$ integrand, also exhibit relatively large exponents ($n\simeq3.57$ and $3.52$, respectively).
The temperature depletion of $D_s$ depends on the excitation of low-energy BdG quasiparticles and their overlap with the momentum-space regions that contribute strongly to the superfluid weight. The fitted exponents also depend on filling due to different relative percentages of the intra- and interband components of $D^s_{\mu\mu}$, which exhibit different temperature dependencies (Fig.~\ref{fig:figem2} in the End Matter), as well as filling dependent changes in the pairing structure. For the nematic states, the broken $C_{3z}$ symmetry additionally leads to an anisotropic temperature dependence, and hence the fitted exponents differ between $D^{s}_{xx}$ and $D^{s}_{yy}$ (see Fig.~S20 of SM~\cite{SM}).

{\it{Discussion.}}---In summary, we have obtained self-consistent superconducting pairings in twisted bilayer graphene within a BCS model of phenomenological attractive interactions in the THF model. Our analysis reveals several competing superconducting states. We find that, although the superconducting order parameter predominantly resides on the heavy electrons, the light electrons are indispensable for the experimentally relevant phase stiffness. Decomposition of superfluid stiffness reveals that apart from the usual interband quantum geometric contribution, even the intraband contribution in TBG is related to quantum geometry due to the topological $f$-$c$ hybridization and quantum geometry concentration near the $\Gamma$-point. An $f$-only model severely underestimates the maximal values of the superfluid weight. The superfluid weight and resulting estimated transition temperatures match recent experiments near the filling $\nu \simeq -2.9$, and the temperature dependence of the stiffness provides information about the pairing symmetry. These results suggest that superconductivity in TBG combines heavy-fermion pairing with light-fermion quantum-geometric supercurrent. An experimental test of our predictions would be provided by a measurement of the superconducting gap and the superfluid weight from the same sample, as a function of filling and temperature. The approach developed here can be extended to other twisted flat-band materials. 

%\begin{acknowledgments}
{\it{Acknowledgments.}}---We thank Tero Heikkil\"a, Guodong Jiang, Junze Deng, Kryštof Kolář, and Zhida Song for helpful discussions. We thank Lorenzo Crippa for sharing data from Ref.~\cite{Rai2023}. This work was supported by a collaboration between The Kavli Foundation, Klaus Tschira Stiftung, and Kevin Wells, and by the Jane and Aatos Erkko Foundation, the Keele Foundation, and the Magnus Ehrnrooth Foundation, as part of the SuperC collaboration.
This work was supported by a grant from the Simons Foundation (SFI-MPS-NFS-00006741-01, B.A.B; SFI-MPS-NFS-00006741-12, P.T.) in the Simons Collaboration on New Frontiers in Superconductivity. B.A.B. was supported by Office of Basic Energy Sciences, Material Sciences and Engineering
Division, U.S. Department of Energy (DOE) under Contracts No. DE-SC0016239.  H.H was also supported by the Gordon and Betty Moore Foundation through Grant No. GBMF8685 towards the Princeton theory program, the Gordon and Betty Moore Foundation's EPiQS Initiative (Grant No. GBMF11070), the Office of Naval Research (ONR Grant No. N00014-20-1-2303), the Global Collaborative Network Grant at Princeton University, the Simons Investigator Grant No. 404513,  the NSF-MERSEC (Grant No. MERSEC DMR 2011750), Princeton Catalysis Initiative, the Schmidt Foundation at the Princeton University, and the National Science Foundation through the AI Research Institutes program Award No. DMR-2433348. B.A.B. was supported by a European Research Council (ERC) under the European Union’s Horizon 2020 research and innovation program (Grant Agreement No. 101020833). 
This work was supported by the Research Council of Finland under project number 374169 and is part of the Finnish Centre of Excellence in Quantum Materials (QMAT).
%\end{acknowledgments}

%TC:ignore

\bibliographystyle{apsrev4-1}
\bibliography{Intro,refmagnetic,tbg_gap,tbg,sm}

\section*{End Matter}

\section*{Model, methods, and parameters.}
We employ the THF model to describe the normal-state properties of TBG~\cite{Song2022}. The model is built on the low-energy wavefunctions of the Bistritzer-MacDonald continuum model~\cite{Bistritzer2011}, so its parameters are set by the microscopic continuum description of TBG; they can also be treated as phenomenological, but we choose them as fixed from the microscopic description. Within this model, the low-energy electronic states are described by two types of electrons per spin and valley: electrons in two localized (heavy) $f$ orbitals, which are centered at the AA-stacked regions of the moiré lattice, and in four dispersive (light) $c$ orbitals. 
The $f$ orbitals carry about $95\%$ of the spectral weight of the flat bands and are therefore the main source of correlation effects, and the light $c$ electrons account for the remaining spectral weight near the $\Gamma$ point of the moiré BZ by hybridizing with the $f$ electrons, which is necessary for describing the topology of the flat bands correctly. The Hamiltonian of the model takes the following form:

\begin{eqnarray}\label{Ham0}
H_{\rm THF} = \sum_{k a a' \eta s} \mathcal{H}^{c, \eta}_{aa'} ({\bf k}) c^{\dagger}_{ka\eta s} c_{ka'\eta s} +  \nonumber \\ \sum_{k\alpha a \eta s} \left( \mathcal{H}^{fc,\eta}_{\alpha a } ({\bf k}) f^{\dagger}_{k\alpha \eta s} c_{ka\eta s} + h.c.\right) -\mu \, \hat{N}\nonumber \\ + H_{U1} + H_{U2} + H_J + H_V + H_W~,
\end{eqnarray}
where $c_{\mathbf{k} a \eta s}$ annihilates a dispersive electron with orbital $a=1,2,3,4$, valley $\eta = \pm$, and spin $s$, and $f_{\mathbf{k} \alpha \eta s}$ annihilates an $f$ electron with orbital $\alpha=1,2$, $\mu$ is the chemical potential, and $\hat{N}$ is the total particle number operator. The free-particle Hamiltonians of the dispersive $c$ electrons are described by $\mathcal{H}^{c,\eta}$, while $\mathcal{H}^{fc,\eta}$ contains the coupling between the $f$ and the $c$ orbitals. The remaining terms are electronic interactions: $H_{U1}$ and $H_{U2}$ are the onsite and NN repulsion between the $f$ electrons, respectively; $H_J$ is a ferromagnetic exchange between $f$ and $c$ electrons, $H_V$ is repulsion between the $c$ electrons, and $H_W$ denotes the repulsive interactions between $f$ and $c$ electrons. Among these, the onsite repulsion $H_{U_1}$ is the largest interaction scale, and we incorporate its effect through a Fermi-liquid quasiparticle renormalization of the $f$ electrons~\cite{SM}, using filling-dependent quasiparticle weight $Z$ obtained in previous dynamical mean-field theory (DMFT) studies~\cite{Rai2023,Datta2023,calderon2025,Song2022}. The remaining interaction terms, $H_{U_2}$, $H_J$, $H_V$, and $H_W$, are treated at the Hartree level. This defines the correlated NS in the absence of superconductivity or other long-range order, which we use as the starting point for introducing effective SC pairing. The explicit forms of the interaction terms are given in Sec.~S1 of the SM~\cite{SM}.

We assume that the lattice local moments are screened and that the normal-state is a Fermi liquid, as suggested by previous DMFT results showing well-defined coherent quasi-particles with a finite quasi-particle weight \cite{Rai2023,Datta2023}. Within this description, the local self‑energy of the $f$ electrons can be expanded as 
$\Sigma_f(\omega) \simeq \Sigma_f(0) + \omega \left(1-\frac{1}{Z}\right)$, yielding an effective renormalized single‑particle Hamiltonian for the NS in the absence of superconductivity or other long‑range order, which reads as
\begin{equation}\label{heff1}
H_{\mathrm{eff}}
=
\sum_{\mathbf{k},\eta,s}
\begin{pmatrix}
\tilde{f}^{\dagger} _{\mathbf{k}\eta s} & c^{\dagger}_{\mathbf{k}\eta s}
\end{pmatrix}
h_{\mathrm{eff}}^{\eta}(\mathbf{k})
\begin{pmatrix}
\tilde{f}_{\mathbf{k}\eta s}\\ c_{\mathbf{k}\eta s}
\end{pmatrix},
\end{equation}
where the renormalized $f$ fermion operator $f_{\mathbf{k}\alpha\eta s}
\approx
\sqrt{Z}\,
\tilde{f}_{\mathbf{k}\alpha\eta s}$, and

\begin{equation}\label{heff2}
h_{\mathrm{eff}}^{\eta}(\mathbf{k})
=
\begin{pmatrix}
Zh_{f}^{\eta}(\mathbf{k}) &
\sqrt{Z}\,h_{fc}^{\eta}(\mathbf{k}) \\
\sqrt{Z}\,h_{fc}^{\dagger\eta}(\mathbf{k}) &
\,h_{c}^{\eta}(\mathbf{k})
\end{pmatrix}.
\end{equation}
where $h_{fc}^{\eta}(\mathbf{k})$, $h_{c}^{\eta}(\mathbf{k})$ and $h_{f}^{\eta}(\mathbf{k})$ contain shifted energy scales associated to Hartree decomposition of the interaction terms~\cite{SM}.

We then consider the SC pairing between the renormalized coherent $f$ quasiparticles. Previous studies have discussed various types of SC pairing channels for TBG, with different microscopic mechanisms and models favoring distinct symmetry classes~\cite{IMartin2018,Julku2020,Annica2022,Yu2023,Fabrizio2022,Wang2024,baldo2026unifyingdescriptioncompetingchiral,Liu2024,Bernevig2025,Isobe2018,You2019,Pangburn2024}. Experimental results do not yet confirm or strongly suggest a clear electronic or phononic or mixed superconducting mechanism~\cite{Gao2026,Stepanov2020}. 
We hence consider both NN and onsite pairing channels in our study as a proxy for different short/longer range possible attractive interactions. %The inclusion of the nearest-neighbor channel is additionally motivated by susceptibility analysis, which gives an effective pairing interaction with a dominant nearest-neighbor component, even in the absence of phonons \cite{Hu2026}.
We use the following density-density pairing interaction between the coherent quasiparticles $H_{pairing} = -Z^2 \sum_{i} \sum_{\alpha\beta\eta\eta'ss'} U_{\alpha\beta}\tilde{n}^{f}_{i\alpha \eta s} \tilde{n}^{f}_{i\beta \eta' s'}- Z^2\sum_{\langle ij\rangle} \sum_{\alpha\beta\eta\eta's s'} V_{\alpha\beta}\tilde{n}^{f}_{i\alpha \eta s} \tilde{n}^{f}_{j\beta \eta' s'}$, where $U_{\alpha\beta}$ and $V_{\alpha\beta}$ are the onsite and NN pairing strengths, respectively, $\tilde{n}^{f}_{i\alpha \eta s} = \tilde{f}^\dagger_{i\alpha \eta s} \tilde{f}_{i\alpha \eta s}$, and the factor $Z^2$ accounting for quasiparticle renormalization
$f_{i\alpha\eta s}\approx\sqrt{Z}\,\tilde f_{i\alpha\eta s}$ of the interaction term, resulting in the mean-field Hamiltonian
\begin{equation} H_{\Delta} = \frac{1}{2}\sum_{k\eta \eta' s s'\alpha\beta}\Delta^{\eta\eta', ss'}_{\alpha\beta}({\bf k}) \tilde{f}_{{\bf k}\alpha \eta s} \tilde{f}_{-{\bf k}\beta\eta's'} + H.c. ~, 
\end{equation}
where $\Delta^{\eta\eta', ss'}_{\alpha\beta}({\bf k})$ encodes the orbital structure and momentum dependence of the SC order parameter, obtained self-consistently by solving the gap equation, see Sec.~S4 of SM~\cite{SM}. We consider NN pairing both for intraorbital ($V_{\alpha \alpha} = V$, $V_{\alpha \beta} = 0$ for $\alpha \neq \beta$) and interorbital ($V_{\alpha \alpha} = 0$, $V_{\alpha \beta} = V$ for $\alpha \neq \beta$) interactions.
 We treat the pairing interaction strength $V$ phenomenologically and choose its value so that the resulting SC gap (maximum value $~ 0.4$ meV between $-3 < \nu < -2$ unless otherwise stated) is comparable to the experimentally observed local gap scale inferred from scanning tunneling microscopy coherence peaks~\cite{Oh2021,Nadjperge2026,pablo2026}.
\section*{Details of the reduced $f$-only model}

To investigate the role of the light $c$ electrons in the superfluid weight, we consider a reduced two-$f$-orbitals of $p_\pm$ character on the triangular lattice formed by the AA regions,

\begin{equation}
H_f=\sum_{\mathbf{k},\eta,s}
\tilde f_{\mathbf{k}\eta s}^{\dagger}
h_f^\eta(\mathbf{k})
\tilde f_{\mathbf{k}\eta s}.
\end{equation}

Although direct $f$-$f$ hopping is neglected in the full THF model, here we include NN interorbital hopping $t_f\simeq0.15~\mathrm{meV}$ in the reduced model, estimated from the overlap of the THF $f$-orbital Wannier functions~\cite{Song2022}. 
Including quasiparticle renormalization, the Bloch Hamiltonian reads

\begin{equation}
h_f^\eta(\mathbf{k})=
\begin{pmatrix}
-\mu & g_\eta(\mathbf{k})\\
g_\eta^*(\mathbf{k}) & -\mu
\end{pmatrix},
\end{equation}
where $\mu$ is the chemical potential and
\begin{align}
g_\eta(\mathbf{k})
=2Z_0 t_f\sum_{\ell=1}^{3}
e^{i\eta\phi_\ell}
\cos(\mathbf{k}\cdot\boldsymbol{\delta}_\ell),
\\\qquad
(\phi_1,\phi_2,\phi_3)=(0,2\pi/3,-2\pi/3).
\end{align}
Here $\boldsymbol{\delta}_\ell$ are
NN bond vectors of the triangular lattice. The normal-state Hamiltonian is symmetric under orbital-exchange $\sigma_x[h_f^\eta(\mathbf{k})]^*\sigma_x=h_f^\eta(\mathbf{k})$. Hence, the $f$ orbital densities are equal, and the Hartree self-energy is proportional to the identity,
\[
\Sigma_{\rm H}
=
\begin{pmatrix}
\Sigma_{{\rm H},1}&0\\
0&\Sigma_{{\rm H},2}
\end{pmatrix}
=\Sigma_{\rm H}^{0}\sigma_0.
\]
Consequently, the Hartree potential only shifts the band energies uniformly, without modifying the bandwidth or eigenvectors. The reduced model therefore lacks the Hartree-induced band deformation arising from the relative shift of the $f$ and $c$ sectors in the full THF model.

We fix the quasiparticle weight to $Z_0=Z(\nu=-2.9)\simeq0.45$, its value at the optimal filling in the full THF model. The resulting bandwidth is comparable to that of the quasiparticle-renormalized THF flat bands at charge neutrality.
%(see Fig.~S17 of the SM~\cite{SM}). 
We consider intraorbital pairing, choosing the pairing interaction to yield $\Delta\simeq0.4~\mathrm{meV}$ near $\nu=-2.9$. We obtain $D_s\simeq0.1~\mathrm{meV}$ and $T_{\mathrm{BKT}}\simeq0.4\mathrm{K}$ (see Fig.~S17 of SM~\cite{SM}). The small stiffness of the $f$-only model highlights the importance of the dispersive $c$ electrons and $f$-$c$ hybridization in enhancing the superfluid weight in the full THF model.

\clearpage
\onecolumngrid

\begin{center}
    \Large\textbf{Supplemental Material}
\end{center}

\let\addcontentsline\origaddcontentsline
\addtocontents{toc}{\protect\setcounter{tocdepth}{2}}
\tableofcontents

\clearpage
%\appendix

\setcounter{section}{0}
\setcounter{subsection}{0}
\setcounter{equation}{0}
\setcounter{figure}{0}
\setcounter{table}{0}
\setcounter{secnumdepth}{2}

\renewcommand{\thesection}{S\arabic{section}}
\renewcommand{\thesubsection}{\thesection.\arabic{subsection}}
\renewcommand{\thesubsubsection}{\thesubsection.\arabic{subsubsection}}

\renewcommand{\theequation}{S\arabic{equation}}
\renewcommand{\thefigure}{S\arabic{figure}}
\renewcommand{\thetable}{S\arabic{table}}

\section{Interacting Topological Heavy Fermion (THF) model: Normal state}{\label{THFhamil}}

In this section, we give details of all the terms in the Topological Heavy Fermion (THF) model Hamiltonian, Eq.~(4) of the main text, including the detailed forms of the interactions. We adapt the THF model proposed in Ref.~\cite{Song2022}. The non-interacting THF model has two types of fermions: dispersing less correlated conduction $c$-electrons and strongly correlated less dispersing $f$-electrons. The corresponding THF non-interacting Hamiltonian is given by  
\begin{align}
&H_0 =  \sum_{\eta s a a'} \sum_{|\mathbf{k}| < k_c} 
H^{c,\eta}_{a a'}(\mathbf{k}) \, c^\dagger_{\mathbf{k} a \eta s} \, c_{\mathbf{k} a' \eta s} 
+ \sum_{\eta s \alpha a} \sum_{|\mathbf{k}| < k_c} 
\left(  H^{fc,\eta}_{ \alpha a}(\mathbf{k}) \, f^{\dagger}_{\mathbf{k} \alpha \eta s}c_{\mathbf{k} a \eta s} \ + \text{H.c.}  \right)-\mu \, \hat{N},
\label{eq:H0}
\end{align}
where $\mu$ is the chemical potential, $\hat{N}$ is the total particle number operator, $c_{\mathbf{k} a \eta s}$ annihilates a conduction electron with orbital $a=1,2,3,4$, valley $\eta = \pm$, and spin $s$, and $f_{\mathbf{k} \alpha \eta s}$ annihilates an $f$ electron with orbital $\alpha=1,2$. Here, 

\begin{equation}
H^{c,\eta}(\mathbf{k}) =
\begin{pmatrix}
0_{2\times2} & v_{\star}(\eta k_x\sigma_0+i k_y\sigma_z)\\
v_{\star}(\eta k_x\sigma_0-i k_y\sigma_z) & M\sigma_x
\end{pmatrix} \label{cHamiltonianMatrix}
\end{equation}

and 

\begin{equation}
H^{fc,\eta}(\mathbf{k}) =
e^{-|\mathbf{k}|^2\lambda^2/2}
\left(\gamma\sigma_0+v_{\star}'(\eta k_x\sigma_x+k_y\sigma_y), 0_{2\times 2}\right).
\end{equation}
The exponential factor in $H^{fc,\eta}_{\alpha a}$ arises from the spatial spread $\lambda \approx 0.3375 a_M$  of the $f$-orbital Wannier functions at twist angle $\theta=1.05^\circ$, where $a_M$ is the moiré lattice constant.  

The interactions in the basis of the THF model are obtained by
projecting the Coulomb interaction of the Bistritzer-MacDonald (BM) continuum model onto the
localized \(f_{\mathbf{k}\alpha\eta s}\) and itinerant
\(c_{\mathbf{k}a\eta s}\) degrees of freedom. This gives
\begin{equation}
 H_I= H_{U_1}+ H_J+ H_{U_2}+ H_V+ H_W.
\label{eq:HI}
\end{equation}
Here, \( H_{U_1}\) is the onsite repulsion between localized \(f\)-electrons given by
\begin{equation}
 H_{U_1}=\frac{U_1}{2}\sum_{\mathbf{R}} :\rho^f_{\mathbf{R}}::\rho^f_{\mathbf{R}}:,~~~
\rho^f_{\mathbf{R}}=
\sum_{\alpha\eta s}
f^\dagger_{\mathbf{R}\alpha\eta s}
f_{\mathbf{R}\alpha\eta s},
\label{eq:HU1}
\end{equation}
where $:\rho^f_{\mathbf{R}}:$ denotes normal ordering of the operator $\rho^f_{\mathbf{R}}$.

The nearest-neighbor \(f\)-electron repulsion
\( H_{U_2}\) is given by
\begin{equation}
\hat H_{U_2}
=
\frac{U_2}{2}
\sum_{\mathbf{\langle RR'\rangle}}
:\rho^f_{\mathbf{R}}:
:\rho^f_{\mathbf{R'}}: ,
\qquad
\rho^f_{\mathbf{R}}
=
\sum_{\alpha\eta s}
f^\dagger_{\mathbf{R}\alpha\eta s}
f_{\mathbf{R}\alpha\eta s}.
\label{eq:HU2}
\end{equation}
The \(c\)-electron repulsion \( H_V\) is given by

\begin{equation}
\hat H_V =
\frac{1}{2N\Omega_M}\sum_{\mathbf q}
V(\mathbf q):\rho^c_{\mathbf q}::\rho^c_{-\mathbf q}:,
\label{eq:HV}
\end{equation}
where $\rho^c_{\mathbf q}=\sum_a\rho^c_{a,\mathbf q}$,
$\rho^c_{a,\mathbf q}
=\sum_{\mathbf k\eta s}
c^\dagger_{\mathbf k+\mathbf q,a\eta s}
c_{\mathbf k a\eta s}$,
$N$ is the number of moir\'e unit cells,
$\Omega_M$ is the moir\'e unit-cell area, and
$V(\mathbf q)$ is the Fourier transform of the
Coulomb interaction, and $V\equiv V(\mathbf q=0)/\Omega_M$, and the density-density repulsion $H_W$ is
\begin{equation}
\hat H_W =
\frac{1}{N}\sum_{\mathbf R,\mathbf q,a}
W_a e^{i\mathbf q\cdot\mathbf R}
:\rho^f_{\mathbf R}::\rho^c_{a,\mathbf q}:,
\label{eq:HW},
\end{equation}
where $W_a$ denotes the coupling to conduction orbital $a$,
with $W_1=W_2$ and $W_3=W_4$. Finally, $H_J$ describes a ferromagnetic exchange coupling between \(U(4)\) moments of the localized $f$-electrons and and the
\(\Gamma_1\oplus\Gamma_2\) itinerant \(c\)-electron sector~\cite{Song2022}.
The total interacting THF Hamiltonian is then
\begin{equation}
H_{\rm THF}=H_0+H_I, 
\label{eq:THFH}
\end{equation}
with $H_0$ given in Eq.~\eqref{eq:H0} and $H_I$ defined by Eqs.~\eqref{eq:HI}-\eqref{eq:HW}. This total $H_{\rm THF}$ is given in Eq.~(4) of the main text.

\section{Treating normal state interactions in the THF model}{\label{THFhamilDMFTandMF}}

In the THF model, the onsite Coulomb interaction $H_{U_1}$ (Eq.~\eqref{eq:HU1}) is the dominant interaction \cite{Datta2023,Rai2023,Zhou2024}, driving the strong correlation effects within the $f$-sector. Hence, the local quantum fluctuations coming from $H_{U_1}$ in these correlated orbitals lead to spectral weight redistribution, which has been studied within dynamical mean field theory (DMFT) \cite{Datta2023,Rai2023,Zhou2024}. We capture the effect of correlation-induced effects of $H_{U_1}$ by renormalizing $f$-electrons using the self energy obtained from previous DMFT calculations. In particular, we include the dynamical correlation effects of $H_{U_1}$ through the quasiparticle weight $Z$ and treat the static effects $H_{U_1}$ within the static Hartree mean-field theory. Other interactions $H_{U_2}$, $H_{J}$, $H_{V}$, and $H_{W}$ are considerably weaker than $H_{U_1}$, and are also treated with Hartree mean field theory.   

\subsection{Hartree mean field}{\label{THFhamilMF}}

Static Hartree mean field theory of $H_{U_1}$, $H_{U_2}$, $H_{J}$, $H_{V}$, and $H_{W}$ leads to density-dependent shifts of the $f$ and $c$-electron energies. The resulting Hamiltonian is given by

\begin{align}
H_{\rm H} ={}&
\sum_{\mathbf{k},\eta,s,\alpha}
\epsilon_f f^\dagger_{\mathbf{k}\alpha\eta s}
f_{\mathbf{k}\alpha\eta s}
+\sum_{\mathbf{k},\eta,s,a,a'}
h^\eta_{c,aa'}(\mathbf{k})
c^\dagger_{\mathbf{k}a\eta s}
c_{\mathbf{k}a'\eta s}
\nonumber\\
&+\sum_{\mathbf{k},\eta,s,\alpha,a}
\left(
h^\eta_{fc,\alpha a}(\mathbf{k})
f^\dagger_{\mathbf{k}\alpha\eta s}
c_{\mathbf{k}a\eta s}
+\mathrm{H.c.}
\right),
\label{eq:Hh}
\end{align}
where
\begin{align}
h^\eta_{c,aa'}(\mathbf{k})
&=\mathcal{H}^{c,\eta}_{aa'}(\mathbf{k})
-(\mu_{c,a}+\mu)\delta_{aa'},\nonumber\\
h^\eta_{fc}(\mathbf{k})
&=\mathcal{H}^{fc,\eta}(\mathbf{k}),
\label{eq:Hartree_blocks}
\end{align}
where $\epsilon_f=-\mu_f-\mu$ is the $f$-electron chemical potential which is shifted by

\begin{equation}
\mu_f=-\frac{7U_1}{8}\nu_{f} - \bigg(\frac{U_1}{2}-8U_{1}\nu^{f}_{0}\bigg)- 6U_{2}\nu_{f}+48\nu^{f}_{0}U_{2}-\sum_{a}W_{a}\nu_{c,a}+ J\frac{\nu^{\prime}_{c,\alpha+2}}{4},
\label{eq:harf}
\end{equation}

and the corresponding shift in the $c$-electron chemical potential is
\begin{equation}
\mu_{c,a}
=
-V\nu_c
-W_1(\nu_f-8\nu_0^f)
+P(\delta_{a,3}+\delta_{a,4}),
\label{eq:harc}
\end{equation}
where
\begin{equation}
P=
-\left(W_3-W_1-\frac{J}{8}\right)
(\nu_f-8\nu_0^f).
\end{equation}
Here,
\begin{equation}
\nu_f=
\frac{1}{N}
\sum_{\mathbf{k},\alpha,\eta,s}
\left\langle
f^\dagger_{\mathbf{k}\alpha\eta s}
f_{\mathbf{k}\alpha\eta s}
\right\rangle
\end{equation}
is the total $f$-electron occupation per moiré unit cell,
and $\nu_0^f=1/2$ is the neutral reference occupation
per flavor, such that $8\nu_0^f=4$.
The normal-ordered $c$-electron densities are
\begin{equation}
\nu_{c,a}
=
\frac{1}{N_M}
\sum_{\mathbf{k},\eta,s}
\left(
\left\langle
c^\dagger_{\mathbf{k}a\eta s}
c_{\mathbf{k}a\eta s}
\right\rangle-\frac12
\right),
\qquad
\nu_c=\sum_{a=1}^{4}\nu_{c,a}.
\end{equation}

\begin{figure*}[h]
\centering
\includegraphics[width=0.45\textwidth]{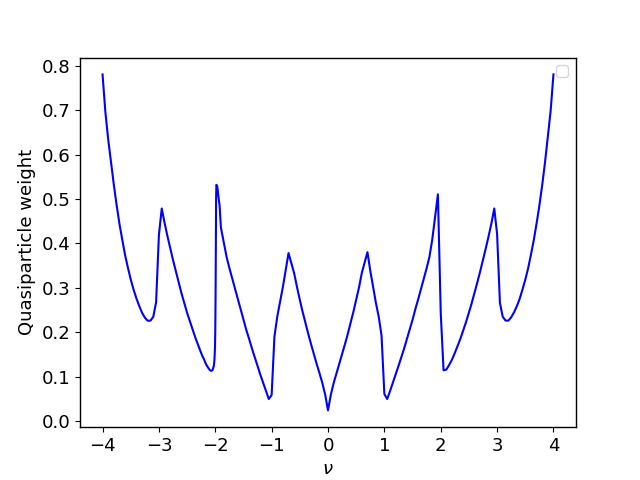}
\caption{The quasiparticle weight $Z$ as a function of filling $\nu$~\cite{Rai2023}.}
\label{fig:figSM0}
\end{figure*}

\subsection{Quasiparticle renormalization}{\label{THFhamilDMFT}}

To incorporate the dynamical correlation effects of $H_{U_1}$, we consider the single particle Green's function $G_{a \alpha  i, a^{\prime}\alpha^{\prime} j}^{\eta s}(\mathbf{k},\tau)=-\langle T_{\tau}\chi_{k a \alpha \eta s,i}(\tau)\chi_{k a^{\prime} \alpha^{\prime} \eta s,j}^{\dagger}(0) \rangle$ in the basis
%\begin{equation}
%\psi_{k a \alpha  \eta s,i}(\tau)=[c_{k a\eta s}(\tau)~~f_{k \alpha \eta s}(\tau)]_i.
%\end{equation}

\begin{equation}
\chi_{\mathbf{k}a\alpha\eta s, i}(\tau)=
\begin{pmatrix}
f_{\mathbf{k}a\alpha\eta s}(\tau)\\
c_{\mathbf{k}a\alpha\eta s}(\tau)
\end{pmatrix}_i,
\end{equation}
where $i,j = 1,..6$. The Green's function follows the Dyson equation
\begin{equation}
\left[G^{\eta s}(\mathbf{k},i\omega_n)\right]^{-1}
=
i\omega_n \mathbb{I}_6
-
H_{\mathrm{H}}^{\eta s}(\mathbf{k})
-
\Sigma^{\eta s}(\mathbf{k},i\omega_n)~,
\label{eq:greensfunc}
\end{equation}
where $H_{H}^{\eta s}(\mathbf{k})$ is the static Hartree mean-field Hamiltonian ($\eta s$ component of $H_{H}$ (Eq.~\eqref{eq:Hh})
and $\Sigma^{\eta s}(\mathbf{k},i\omega_n)$ is the dynamical self-energy. Since in the previous DMFT calculations \cite{Datta2023,Rai2023,Zhou2024} the self-energy is considered local and only affecting the heavy $f$-electrons, we can write $[\Sigma^{\eta s}(\mathbf{k},i\omega_n)]_{i a \alpha,j a^{\prime} \alpha^{\prime}}=\delta_{i a \alpha,j a^{\prime} \alpha^{\prime}}(\delta_{i,1}+\delta_{i,2})\Sigma_f(i\omega_n)$. At low temperatures and in an effective Fermi liquid description, $\Sigma_f(i\omega_n)$ is expanded for low-frequencies as
\begin{equation}
\operatorname{Re}\Sigma_f(\omega+i0^+)
=
\left(1-\frac{1}{Z}\right)\omega
+
D\omega^2
+\cdots ,
\label{eq:resigmaf}
\end{equation}

\begin{equation}
\operatorname{Im}\Sigma_f(\omega+i0^+)
=
-\Gamma\omega^2
+\cdots .
\label{eq:imagsigmaf}
\end{equation}
Here $\epsilon_f$ is the static energy shift of the $f$-electrons obtained using Eq.~\eqref{eq:harf}; $Z$ is the quasiparticle weight, $\Gamma$ characterizes the damping of quasiparticles, and $D$ is a constant. 
Retaining up to the linear order in $\omega$, the inverse Green's function can be written as

\begin{eqnarray}\label{invgf0}
[G^{\eta s}(\mathbf{k},\omega)]^{-1} \simeq \begin{pmatrix}
\left[\omega - \epsilon_f - \left(1-\frac{1}{Z}\right)\omega\right]\mathbb{I}_2
&
-h^{\eta}_{fc}(\mathbf{k})
\\
-h^{\eta\dagger}_{fc}(\mathbf{k})
&
\omega - h^{\eta}_c(\mathbf{k})
\end{pmatrix},
\end{eqnarray}

where the elements of $\tilde{H}^{c,\eta}(\mathbf{k})$ and $\tilde{H}^{fc,\eta}(\mathbf{k})$ are obtained from the static mean field Hamiltonian Eq.~\eqref{eq:Hh}.
This $G^{\eta s}(\mathbf{k},\omega)$ can be written in terms of an effective Hamiltonian $\tilde{H}^{\eta s}_{\rm{eff}}$ as
\begin{eqnarray}\label{invgf1}
[G^{\eta s}(\mathbf{k},\omega)]^{-1} = \hat{Z}^{-1/2}\Big[(\omega + i0^+)\mathbb{I}_6 - \sqrt{\hat{Z}}\tilde{H}^{\eta s}_{H}(\mathbf{k})\sqrt{\hat{Z}} \Big]\hat{Z}^{-1/2} .
\end{eqnarray}
where
\begin{eqnarray}\label{qpmatrix}
\hat{Z} =  \begin{pmatrix}
Z\mathbb{I}_2 & 0 \\ 0 & \mathbb{I}_4
\end{pmatrix}.
\end{eqnarray}
Eqs.~\eqref{invgf1}-\eqref{qpmatrix} define the quasiparticle Hamiltonian 
\begin{equation}\label{nsham}
H^{\eta s}_{\rm eff}(\mathbf{k}) = \sqrt{\hat{Z}}\tilde{H}^{\eta s}_{\rm{H}}(\mathbf{k})\sqrt{\hat{Z}},
\end{equation}
with the quasiparticle Green's function
\begin{equation}
G^{\eta s}_{qp}(\omega,\mathbf{k}) = \Big[(\omega + i0^+)\mathbb{I}_6 - H^{\eta s}_{\rm eff}(\mathbf{k}) \Big]^{-1},
\end{equation}
which is related to the Green's function in Eq.~\eqref{invgf0} as
\begin{equation}
G^{\eta s}(\omega,\mathbf{k})= \sqrt{\hat{Z}}G^{\eta s}_{qp}(\omega,\mathbf{k})\sqrt{\hat{Z}}. 
\end{equation}
Hence, we can relate the renormalized quasiparticles $\tilde{\chi}_{k a \alpha  \eta s,i}$ with the bare quasiparticles as $\chi_{k a \alpha  \eta s,i}=\sqrt{Z}\tilde{\psi}_{k a \alpha  \eta s,i}$ for $i=1,2$ and $\psi_{k a \alpha  \eta s,i}=\tilde{\psi}_{k a \alpha  \eta s,i}$ otherwise, which signifies that the renormalized $f$-fermion operator $\tilde{f}_{\mathbf{k}\alpha\eta s}$ is also related by $f_{\mathbf{k}\alpha\eta s}
=
\sqrt{Z}\,
\tilde{f}_{\mathbf{k}\alpha\eta s}$. 

\section{Obtaining normal state and parameters}\label{sec:method}

\begin{figure*}[h]
\centering
    \includegraphics[width=0.6\textwidth]{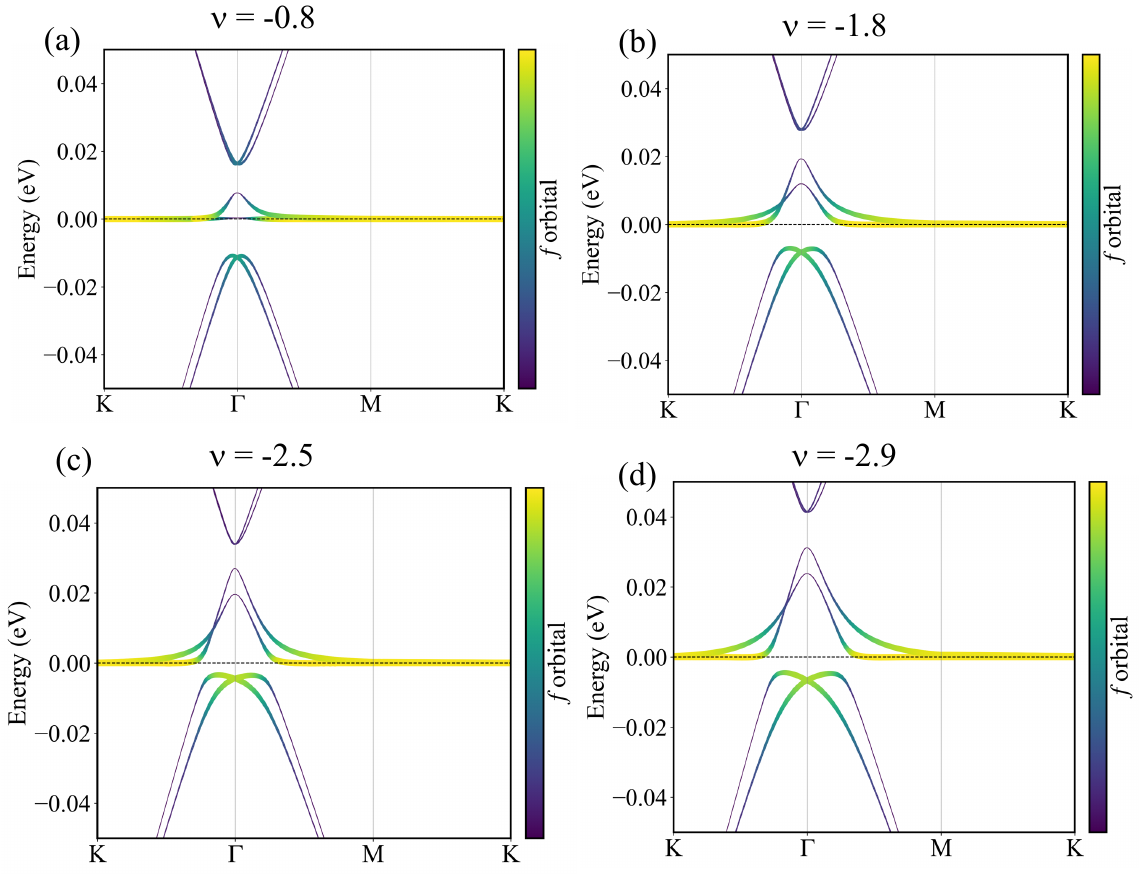}
\caption{Renormalized quasiparticle energy bands at different filling factors with $Z$ from Ref.~\cite{Rai2023}.}
\label{fig:figSM2}
\end{figure*}

We first perform a Hartree decomposition of the interactions to obtain the static energy shifts in Eq.~\eqref{eq:Hh}, and then incorporate the quasiparticle weight $Z$ from previous DMFT studies~\cite{Rai2023}, see Fig.~\ref{fig:figSM0}. With these renormalizations, we then adjust the chemical potential to match the target normal‑state density. The resulting renormalized band structure exhibits a flattened bandwidth and reduced $f$-$c$ hybridization compared to the $Z=1$ case, as shown in Fig.~\ref{fig:figSM2}. Near $\nu \sim 0$, the correlation effects and band renormalization are extremely strong, and our renormalized quasiparticle approach is likely to fail there; however, this does not pose a problem because at those fillings the quasiparticle weight goes to zero and superconductivity vanishes anyway. In contrast, for the fillings where our model predicts superconductivity, our renormalized quasiparticle bands are similar to those obtained directly by DMFT.
We use the THF interaction parameters \(U_1=57.95\,\mathrm{meV}\), \(V=48.33\,\mathrm{meV}\), \(W_1=W_2=44.03\,\mathrm{meV}\), \(W_3=W_4=50.20\,\mathrm{meV}\),  \(J=16.38\,\mathrm{meV}\), and \(U_2=6\,\mathrm{meV}\)~\cite{Song2022}.

\section{BCS superconductivity in the THF model}
\label{THFSChamil}

As discussed in the main text, we describe the correlated normal state
within an effective Fermi-liquid approximation. The physical $f$-electrons and the coherent quasiparticles are related by
\begin{equation}
f_{i\alpha\eta s}
\simeq
\sqrt{Z}\,
\widetilde f_{i\alpha\eta s},
\end{equation}
such that the $f$-$c$ hybridization and the $f$-sector of the
quasiparticle Hamiltonian are renormalized by $\sqrt{Z}$ and $Z$,
respectively.

Since the microscopic origin of superconductivity in TBG remains
unsettled, we introduce phenomenological attractive interactions
directly between the coherent $f$ quasiparticles. We consider onsite
and nearest-neighbor (NN) density-density interactions,
\begin{align}
H_{pairing}
={}&
-U_{\alpha\beta}Z^2
\sum_i
\sum_{\alpha\beta\eta\eta'ss'}
\tilde{n}^{f}_{i\alpha\eta s}
\tilde{n}^{f}_{i\beta\eta's'}
\nonumber\\
&-
V_{\alpha\beta}Z^2
\sum_{\langle ij\rangle}
\sum_{\alpha\beta\eta\eta'ss'}
\tilde{n}^{f}_{i\alpha\eta s}
\tilde{n}^{f}_{j\beta\eta's'},
\label{eq:pairinteraction}
\end{align}
where
\begin{equation}
\tilde{n}^{f}_{i\alpha\eta s}
=
\widetilde f^\dagger_{i\alpha\eta s}
\widetilde f_{i\alpha\eta s},
\end{equation}
and $\langle ij\rangle$ denotes NN sites of the moir\'e triangular
lattice. The interaction strengths $U_{\alpha\beta}$ and $V_{\alpha\beta}$ are phenomenological parameters for onsite and nearest-neighbor (NN) pairing, respectively. For each pairing channel, the attractive interaction is restricted to the selected intraorbital or interorbital components. The interaction strengths are chosen such that the resulting superconducting gap is comparable to the experimentally relevant gap scale. The resulting pairing term is

\begin{equation}\label{pairingham}
H_{\Delta}
=
\frac{1}{2}\sum_{\mathbf{k},\alpha,\beta,\eta,\eta',s,s'}
\Delta_{\alpha\beta}^{\eta\eta', ss'}(\mathbf{k})
\tilde f_{\mathbf{k}\alpha\eta s}
\tilde f_{-\mathbf{k}\beta\eta' s'}
+\mathrm{H.c.}
\end{equation}
Here $\Delta_{\alpha\beta}^{\eta\eta', ss'}(\mathbf{k})$ are the matrix
elements of the pairing order parameter (OP) $\Delta(\mathbf{k})$
in orbital-valley space,
\begin{equation}
\left[\Delta(\mathbf{k})\right]_{\alpha\eta s,\beta\eta's'}
=
\Delta_{\alpha\beta}^{\eta\eta', ss'}(\mathbf{k}),
\end{equation}
which is determined self-consistently.

We focus on intervalley pairing, $\eta'=-\eta$, as discussed in the main text, since it pairs states at opposite momenta that are degenerate in the normal state. Due to the independent spin-rotation symmetry in the two valleys, described by $U(2)_+\times U(2)_-$, intervalley spin-singlet and spin-triplet
pairing states are symmetry-related and degenerate. For example, consider the transformation
\begin{equation}
g f^\dagger_{\mathbf{k}\alpha+\uparrow} g^{-1}
=
f^\dagger_{\mathbf{k}\alpha+\uparrow},
\qquad
g f^\dagger_{\mathbf{k}\alpha+\downarrow} g^{-1}
=
-f^\dagger_{\mathbf{k}\alpha+\downarrow},
\end{equation}
while the operators in the - valley remain unchanged,
\begin{equation}
g f^\dagger_{\mathbf{k}\alpha-s} g^{-1}
=
f^\dagger_{\mathbf{k}\alpha-s}.
\end{equation}
Under this transformation, the intervalley spin-singlet pair transforms as
\begin{align}
g\Big[
&f^\dagger_{\mathbf{k}\alpha+\uparrow}
f^\dagger_{-\mathbf{k}\alpha'-\downarrow}
-
f^\dagger_{\mathbf{k}\alpha+\downarrow}
f^\dagger_{-\mathbf{k}\alpha'-\uparrow}
\Big]g^{-1}
\nonumber\\
&=
f^\dagger_{\mathbf{k}\alpha+\uparrow}
f^\dagger_{-\mathbf{k}\alpha'-\downarrow}
+
f^\dagger_{\mathbf{k}\alpha+\downarrow}
f^\dagger_{-\mathbf{k}\alpha'-\uparrow},
\end{align}
which is the $S_z=0$ spin-triplet pair. Thus, an allowed relative
spin rotation between the two valleys relates the intervalley
spin-singlet and spin-triplet sectors. We therefore choose the
spin-singlet representative for the symmetry classification below;
the corresponding spin-triplet partners are obtained by
$U(2)_+\times U(2)_-$ rotations. 

\subsection{Bogoliubov--de Gennes Hamiltonian}

We now write the BdG Hamiltonian $H_{\rm BdG} = H_{\rm eff} + H_\Delta$ in the Nambu basis, with $H_{\rm eff}$ and $H_\Delta$ defined in Eqs.~\eqref{nsham} and \eqref{pairingham}, respectively.

\begin{equation}
\Psi_{\mathbf{k},\eta}
=
\begin{pmatrix}
\widetilde f_{\mathbf{k}\eta \uparrow}\\
c_{\mathbf{k}\eta \uparrow}\\
\widetilde f^\dagger_{-\mathbf{k}\bar\eta \downarrow}\\
c^\dagger_{-\mathbf{k}\bar\eta \downarrow}
\end{pmatrix}.
\label{eq:Nambu}
\end{equation}
Here
\begin{equation}
\widetilde f_{\mathbf{k}\eta s}
=
\begin{pmatrix}
\widetilde f_{\mathbf{k}1\eta s}\\
\widetilde f_{\mathbf{k}2\eta s}
\end{pmatrix},
\qquad
c_{\mathbf{k}\eta s}
=
\begin{pmatrix}
c_{\mathbf{k}1\eta s}\\
c_{\mathbf{k}2\eta s}\\
c_{\mathbf{k}3\eta s}\\
c_{\mathbf{k}4\eta s}
\end{pmatrix}.
\end{equation}
The BdG Hamiltonian is
\begin{equation}
H_{\rm BdG}
=
\sum_{\mathbf{k},\eta}
\Psi^\dagger_{\mathbf{k},\eta}
\mathcal H_{\rm BdG}^{\eta\bar{\eta}}(\mathbf{k})
\Psi_{\mathbf{k},\eta}.
\label{eq:Hbdg}
\end{equation}
Since the attractive interaction acts only in the $f$-sector, the
pairing matrix is nonzero only in the $f$-orbital block. 
%For the $(+,\uparrow)$--$(-,\downarrow)$ sector,
\begin{equation}
\mathcal{H}^{\eta\bar{\eta}}_{\mathrm{BdG}}(\mathbf{k})
=
\begin{pmatrix}
Zh^\eta_f(\mathbf{k})
&
\sqrt{Z}\,h^\eta_{fc}(\mathbf{k})
&
\Delta^{\eta\bar{\eta}}(\mathbf{k})
&
0
\\
\sqrt{Z}\,h^{\dagger\eta}_{fc}(\mathbf{k})
&
h^{\eta}_c(\mathbf{k})
&
0
&
0
\\
\Delta^{\eta\bar{\eta}\dagger}(\mathbf{k})
&
0
&
-Z[h^{\bar{\eta}}_f(\mathbf{-k})]^*
&
-\sqrt{Z}\,[h^{\bar\eta}_{fc}(\mathbf{-k})]^*
\\
0
&
0
&
-\sqrt{Z}\,[h^{\bar\eta}_{fc}(\mathbf{-k})]^T
&
-[h^{\bar\eta}_c(\mathbf{-k})]^*
\end{pmatrix}.
\label{eq:BdGmatrix}
\end{equation}
Here $\Delta^{\eta\bar{\eta}}(\mathbf{k})$ is the $2\times2$ pairing matrix in the
$f$-orbital subspace. The other intervalley sector of the BdG Hamiltonian has the identical quasiparticle spectrum.

The BdG eigenvalue equation is
\begin{equation}
\mathcal H_{\rm BdG}^{\eta\bar{\eta}}(\mathbf{k})
\psi_{\mathbf{k}n\eta s}
=
E_{\mathbf{k}n\eta}
\psi_{\mathbf{k}n\eta s},
\label{eq:BdGeigen}
\end{equation}
where the eigenvector contains the particle and hole components,
\begin{equation}
\psi_{\mathbf{k}n\eta}
=
\begin{pmatrix}
u_{\mathbf{k}n\eta}\\
v_{\mathbf{k}n\eta}
\end{pmatrix}.
\end{equation}
Using these eigenvalues and eigenvectors, we then solve the gap equations self-consistently.

\subsection{Self-consistent pairing order parameter}\label{gapeqn}

For the NN attractive interaction, the pairing amplitude on
the bond $\boldsymbol{\delta}_{\ell}$ is determined self-consistently
from
\begin{equation}
\Delta_{\alpha\beta}^{\ell}
=
-\frac{Z^2V_{\alpha\beta}}{N}
\sum_{\mathbf{k}}
e^{-i\mathbf{k}\cdot\boldsymbol{\delta}_{\ell}}
\left\langle
\widetilde f_{-\mathbf{k}\beta-\downarrow}
\widetilde f_{\mathbf{k}\alpha+\uparrow}
\right\rangle.
\label{eq:scmf}
\end{equation}
The corresponding momentum-dependent pairing matrix
entering the BdG Hamiltonian is
\begin{equation}
\Delta_{\alpha\beta}^{+-}(\mathbf{k})
=
\sum_{\ell}
\Delta_{\alpha\beta}^{\ell}
e^{i\mathbf{k}\cdot\boldsymbol{\delta}_{\ell}}.
\label{eq:phik}
\end{equation}
For onsite pairing, the effective quasiparticle coupling
is $Z^2U_{\alpha\beta}$, and the self-consistency equation
contains no bond form factor. We solve the self-consistent gap equations and compare the grand potentials of the resulting solutions at fixed normal-state chemical potential. For NN pairing, the grand potential per moiré unit cell is given by

\begin{align}
\Omega
={}&
-\frac{1}{N\beta}
\sum_{\mathbf{k},\eta,n}
\ln\left(
1+e^{-\beta E_{\mathbf{k}n\eta}}
\right)
+
\sum_{\alpha\beta,\ell}
\frac{
|\Delta_{\alpha\beta}^{\ell}|^2
}{Z^2 V_{\alpha\beta}} 
+\mathrm{const.}
\label{eq:grandpotential}
\end{align}
When multiple solutions are obtained, their grand potentials are compared at fixed chemical potentials, and the
solution with the lowest $\Omega$ is identified as the lowest-grand-potential
state.

\subsection{Symmetry transformations of the pairing order parameter}

We next provide the explicit symmetry transformations used for the
classification of the self-consistent pairing states discussed in the
main text. The normal-state THF Hamiltonian is invariant under time
reversal ($\mathcal{T}$) and the $D_6$ point group generated by $C_{3z}$, $C_{2z}$,
and $C_{2x}$. 
Their actions on the $f$-electron creation operators are
\begin{align}
\mathcal{T}
\tilde{f}^\dagger_{\mathbf{k}\alpha\eta s}
\mathcal{T}^{-1}
&=
s\tilde{f}^\dagger_{-\mathbf{k}{\alpha}\bar{\eta}-s},
\nonumber\\
C_{2z}
\tilde{f}^\dagger_{\mathbf{k}\alpha\eta s}
C_{2z}^{-1}
&=
\tilde{f}^\dagger_{-\mathbf{k}\bar{\alpha}\bar{\eta}s},
\nonumber\\
C_{3z}
\tilde{f}^\dagger_{\mathbf{k}\alpha\eta s}
C_{3z}^{-1}
&=
e^{i\frac{2\pi}{3}\eta(-1)^{\alpha-1}}
\tilde{f}^\dagger_{C_{3z}\mathbf{k},\alpha\eta s},
\nonumber\\
C_{2x}
\tilde{f}^\dagger_{\mathbf{k}\alpha\eta s}
C_{2x}^{-1}
&=
\tilde{f}^\dagger_{C_{2x}\mathbf{k}\bar{\alpha}\eta s},
\label{eq:symmetryaction}
\end{align}
where $\bar{\alpha}=3-\alpha$ and $\bar{\eta}=-\eta$.

Considering the first operator part of $H_\Delta$ in Eq.~\eqref{pairingham}, for intervalley pairing, the pairing operator is of the form
\begin{equation}
\hat{\Delta}
=
\sum_{\mathbf{k},\alpha\alpha',ss'}
\Delta_{\alpha\alpha'}^{+-,ss'}(\mathbf{k})
\tilde f^\dagger_{\mathbf{k}\alpha+s}
\tilde f^\dagger_{-\mathbf{k}\alpha'-s'},
\label{eq:pairing_op}
\end{equation}
the action of the threefold rotation is
\begin{align}
C_{3z}\hat{\Delta}C_{3z}^{-1}
={}&
\sum_{\mathbf{k},\alpha\alpha',ss'}
e^{i\frac{2\pi}{3}
\left[(-1)^{\alpha-1}-(-1)^{\alpha'-1}\right]}
\Delta_{\alpha\alpha'}^{+-,ss'}
(C_{3z}^{-1}\mathbf{k})
\nonumber\\
&\times
\tilde f^\dagger_{\mathbf{k}\alpha+ s}
\tilde f^\dagger_{-\mathbf{k}\alpha'-s'}.
\label{eq:C3z_pairing}
\end{align}
The twofold rotations act as
\begin{align}
C_{2z}\hat{\Delta}C_{2z}^{-1}
={}&-
\sum_{\mathbf{k},\alpha\alpha',ss'}
\Delta_{(3-\alpha')(3-\alpha)}^{+-,s's}
(\mathbf{k})
\nonumber\\
&\times
\tilde f^\dagger_{\mathbf{k}\alpha + s}
\tilde f^\dagger_{-\mathbf{k}\alpha'-s'},
\label{eq:C2z_pairing}
\end{align}
\begin{align}
C_{2x}\hat{\Delta}C_{2x}^{-1}
={}&
\sum_{\mathbf{k},\alpha\alpha',ss'}
\Delta_{(3-\alpha)(3-\alpha')}^{+-,ss'}
(C_{2x}^{-1}\mathbf{k})
\nonumber\\
&\times
\tilde f^\dagger_{\mathbf{k}\alpha+ s}
\tilde f^\dagger_{-\mathbf{k}\alpha'-s'},
\label{eq:C2x_pairing}
\end{align}
and $\mathcal{T}$ gives
\begin{align}
\mathcal{T}\hat{\Delta}\mathcal{T}^{-1}
={}&-
\sum_{\mathbf{k},\alpha\alpha',ss'}
ss'
\left[
\Delta_{\alpha'\alpha}^{+-, -s',-s}
(\mathbf{k})
\right]^*
\nonumber\\
&\times
\tilde f^\dagger_{\mathbf{k}\alpha+s}
\tilde f^\dagger_{-\mathbf{k}\alpha'-s'}.
\label{eq:TR_pairing}
\end{align}

Applying Eqs.~\eqref{eq:C3z_pairing}–\eqref{eq:TR_pairing} to the self-consistent pairing matrix we identify the symmetry transformations of the SC solutions and classify them according to the irreducible representations of $D_6$ point group, as summarized in Table I of the main text.

\section{Symmetry classification of mean-field pairing states}
\label{sec:SM_SCstates}
We now apply this classification to the self-consistent solutions obtained for inter and intra-$f$ orbital pairing interactions. 

\subsection{Interorbital nearest-neighbor pairing}

\begin{table}[t]
\centering
\begin{tabular}{c|rrr}
$D_3$ irrep & $1$ & $2C_3$ & $3C_2'$ \\
\hline
$A_1$ & 1 &  1 &  1 \\
$A_2$ & 1 &  1 & -1 \\
$E$   & 2 & -1 &  0
\end{tabular}
\caption{Character table of $D_3$.}
\label{tab:D3}
\end{table}
We first consider interorbital NN pairing at fillings $\nu\approx-2.9$ and $\nu\approx-1.9$ where the SC gap is maximal within $-3< \nu <-2$ and $-2< \nu <-1$, respectively. 
The self-consistent mean-field calculations yield several closely competing superconducting solutions, summarized in Table~I of the main text. As shown in Eq.~\eqref{eq:C2z_pairing}, the intervalley interorbital spin-singlet (triplet) gap is $C_{2z}$ even (odd), which restricts the spin-singlet (triplet) solutions to the $A_1\oplus A_2\oplus E_2$ ($B_1\oplus B_2\oplus E_1$) representations. Here we focus on the spin-singlet sector, for which, the representation classification is equivalent to that of the group $D_6/\{1,C_{2z}\}\cong D_3$ (character table in Table~\ref{tab:D3}). The two lowest-energy states, SC1 and SC2, both preserve $C_{2z}$ and break the $C_{3z}$ and $\mathcal{T}$. SC1 additionally preserves $C_{2x}$ with eigenvalue $+1$, whereas SC2 breaks this symmetry. Both SC1 and SC2 belong to the $A_1\oplus E_2$ representation. 

Projecting the pairing amplitude onto the irreducible representations using $P_X=\frac{d_X}{|G|}\sum_{g\in G}\chi_X(g)^*\hat g$, we extract the weight of each symmetry component. Here, $G=D_3$, $|G|=6$, $d_X$ is the irrep dimension, and $\chi_X(g)$ is the character for the group element $g$. Near $\nu=-2.9$, SC1 contains approximately ($\sim49\%$) $A_1$ and ($\sim51\%$) $E_2$ character, while SC2 contains ($\sim 44\%$) $A_1$ and ($\sim 55\%$) $E_2$ character. The momentum dependence of the lowest harmonics from NN pairing contains admixture of even and odd NN harmonics. 

The even-momentum contribution is approximately $12\%$ for SC1 and $\sim8\%$ for SC2 near $\nu = -2.9$, $\sim10\%$ for SC1 and SC2 near $\nu = -1.9$. The even and odd-momentum-parity-weights are obtained by
decomposing the NN bond amplitudes as $\Delta_{\alpha\beta,\ell}^{e/o}
=
\frac{1}{2}
\left(
\Delta_{\alpha\beta}^{\ell}
\pm
\Delta_{\alpha\beta}^{\bar{\ell}}
\right)$, where $\bar{\ell}$ denotes the bond opposite to
$\ell$. The corresponding normalized weights are $w_{e/o}
=
\frac{
\sum_{\alpha\beta,\ell}
|\Delta_{\alpha\beta,\ell}^{e/o}|^2
}{
\sum_{\alpha\beta,\ell}
\left(
|\Delta_{\alpha\beta,\ell}^{e}|^2
+
|\Delta_{\alpha\beta,\ell}^{o}|^2
\right)
}$. 

Considering $\sigma_i$, $\tau_i$, and $s_i$ denote Pauli
matrices in orbital, valley, and spin space, respectively, in the interorbital, intervalley sector, the orbital-valley
matrices form two doublets,
\begin{align}
\mathcal M_+
&=
\left\{
\sigma_x\otimes\tau_x,\,
(i\sigma_y)\otimes(i\tau_y)
\right\},\\
\mathcal M_-
&=
\left\{
\sigma_x\otimes(i\tau_y),\,
(i\sigma_y)\otimes\tau_x
\right\},
\end{align}
which transform according to the $E_2$ and $E_1$ irreducible
representations of $D_6$, respectively. The fermionic antisymmetry requires
\begin{equation}
\Delta(\mathbf k)=-\Delta^T(-\mathbf k),
\end{equation}
where the transpose acts on orbital, valley, and spin indices.
For a spin-singlet gap,
\begin{equation}
\Delta(\mathbf k)=\Phi_s(\mathbf k)\otimes i s_y,
\end{equation}
the antisymmetry of the spin matrix implies
\begin{equation}
\Phi_s(\mathbf k)=\Phi_s^T(-\mathbf k).
\end{equation}
Consequently, even momentum harmonics multiply the symmetric doublet
$\mathcal M_+$, whereas odd momentum harmonics multiply the
antisymmetric doublet $\mathcal M_-$. Writing $M_{\pm,\ell}$, with
$\ell=1,2$, for the two doublets, the singlet gap-function can be expressed as a sum of factorized momentum-orbital-valley components as

\begin{equation}
\Delta(\mathbf k)
=
\left[
\sum_{\ell=1}^{2}
\phi^{e}_{s,\ell}(\mathbf k)M_{+,\ell}
+
\sum_{\ell=1}^{2}
\phi^{o}_{s,\ell}(\mathbf k)M_{-,\ell}
\right]\otimes i s_y,
\end{equation}
where
\begin{equation}
\phi^{e}_{s,\ell}(-\mathbf k)
=
\phi^{e}_{s,\ell}(\mathbf k),
\qquad
\phi^{o}_{s,\ell}(-\mathbf k)
=
-\phi^{o}_{s,\ell}(\mathbf k).
\end{equation}

For spin-triplet pairing, the spin matrices $s_a i s_y$, with
$a=x,y,z$, are symmetric. Thus,
\begin{equation}
\Delta_t(\mathbf k)
=
\sum_{a=x,y,z}
\Phi_{t,a}(\mathbf k)\otimes(s_a i s_y)
\end{equation}
requires
\begin{equation}
\Phi_{t,a}(\mathbf k)
=
-\Phi_{t,a}^T(-\mathbf k).
\end{equation}
The orbital-valley matrix assignments are therefore reversed: 

\begin{equation}
\Delta(\mathbf k)
=
\sum_{a=x,y,z}
\left[
\sum_{\ell=1}^{2}
\phi^{e}_{t,a\ell}(\mathbf k)M_{-,\ell}
+
\sum_{\ell=1}^{2}
\phi^{o}_{t,a\ell}(\mathbf k)M_{+,\ell}
\right]\otimes(s_a i s_y).
\end{equation}

\begin{figure*}[h]
\centering
\includegraphics[width=0.9\textwidth]{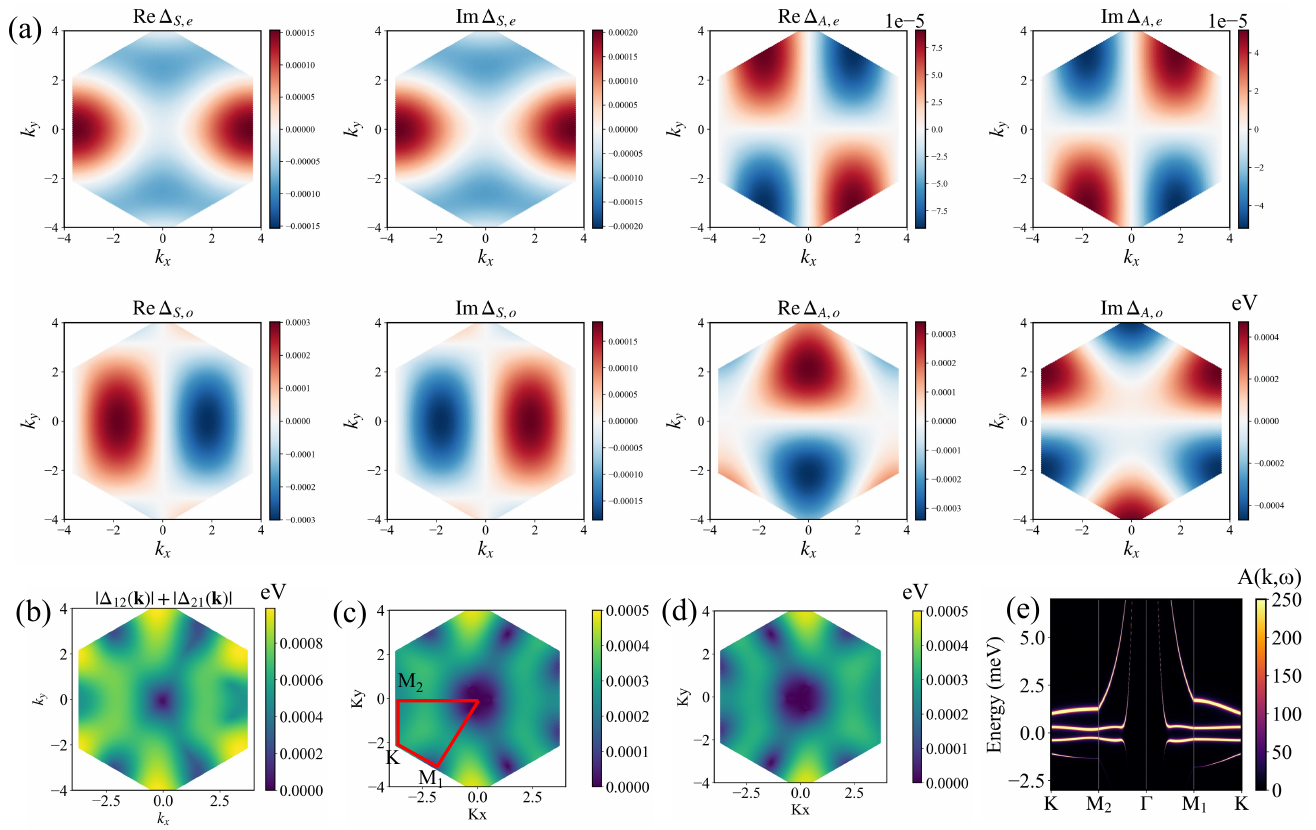}
\caption{Interorbital NN pairing: orbital and momentum-parity decomposition of SC1 pairing. (a) Real and imaginary parts of $\Delta_{S,e}$
, $\Delta_{A,e}$ (upper row) and $\Delta_{S,o}$, $\Delta_{A,o}$
(lower row), with $\Delta_{S,A}(\mathbf{k})= (\Delta_{12}(\mathbf{k}) \pm \Delta_{21}(\mathbf{k}))/2$ and
the superscripts denoting even/odd momentum parity. (b)
$|\Delta_{12}(\mathbf{k})| + |\Delta_{21}(\mathbf{k})|$ in the orbital basis. (c,d) Magnitudes of band-projected gaps on the Fermi surface for the pairing from the two inter-valley BdG blocks. (e) $f$-electron spectral function $A_f(\mathbf{k},\omega)
=-\frac{Z_f}{\pi}\operatorname{Im}
\operatorname{Tr}\left[
P_f^e G^R(\mathbf{k},\omega)
\right]$ using $\eta=0.01$meV, 
along a high-symmetry path 
$K$ - $M_2$ - $\Gamma$ - $M_1$ - $K$, 
indicated by the red path in (c).}
\label{fig:SC1pairing}
\end{figure*}

\begin{figure*}[h]
\centering
    \includegraphics[width=0.9\textwidth]{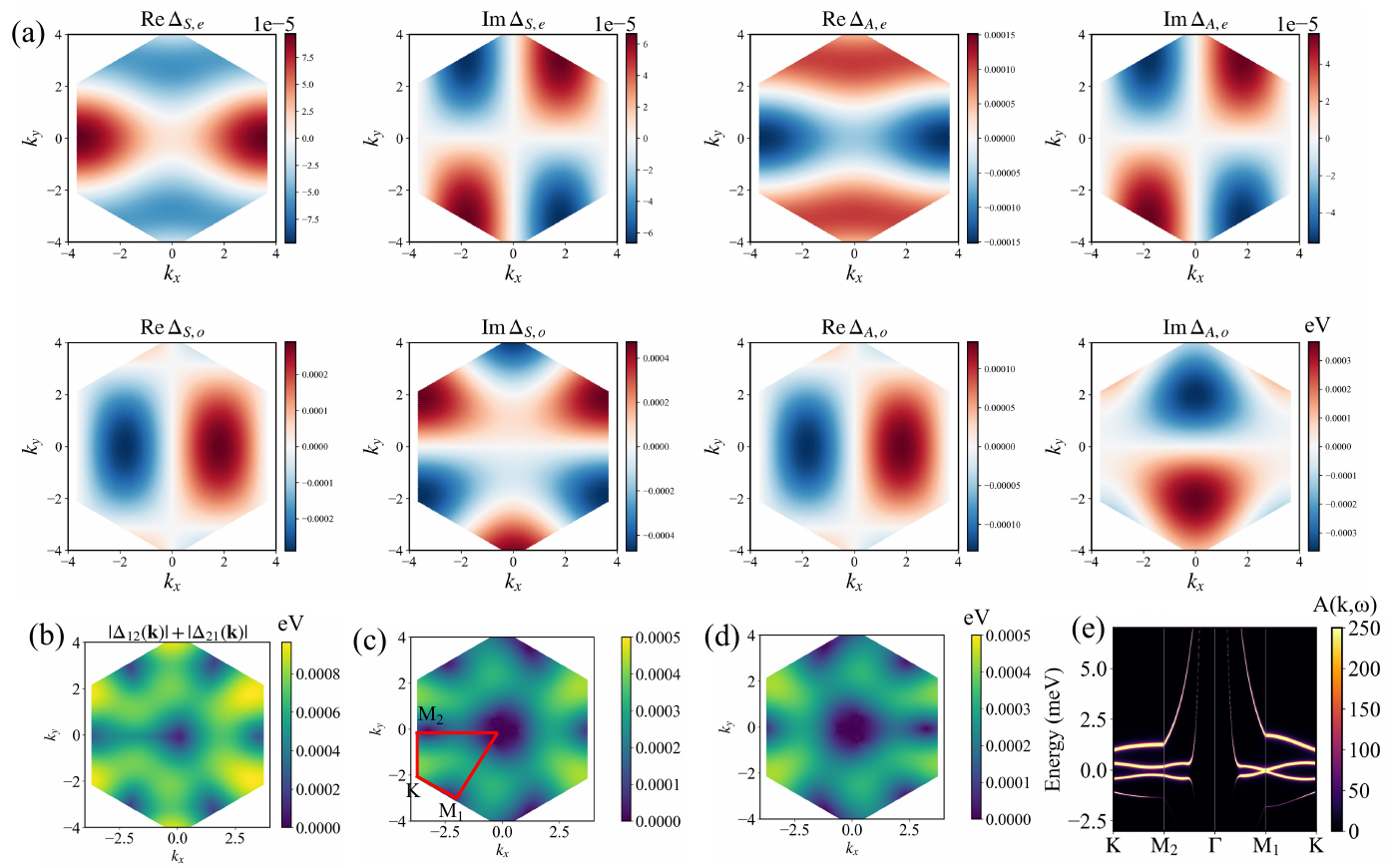}
\caption{Same as Fig.~\ref{fig:SC1pairing}, but for the SC2 pairing state.}
\label{fig:SC2pairing}
\end{figure*}

The NN momentum harmonics comprise extended $s$-wave ($A_1$),
$d$-wave ($E_2$), $p$-wave ($E_1$), and $f$-wave
($B_2$) components. The momentum $\otimes$ orbital-valley components of our numerically obtained solutions (Fig.~\ref{fig:SC1pairing}, \ref{fig:SC2pairing}) transform as $A_1\otimes E_2$, $E_2\otimes E_2$,
$E_1\otimes E_1$, and $B_2\otimes E_1$ for the spin-singlet case. In each of these products, the momentum and orbital-valley factors have
the same $C_{2z}$ parity, so every component lies in the $C_{2z}=+1$
sector. For the spin-triplet case, the corresponding products are
$A_1\otimes E_1$, $E_2\otimes E_1$, $E_1\otimes E_2$, and
$B_2\otimes E_2$, respectively. Here, the momentum and orbital-valley
factors have opposite $C_{2z}$ parity, so all components belong to the
$C_{2z}=-1$ sector.

In addition to SC1 and SC2, we obtain self-consistent solutions, SC3-SC6, which preserve the $C_{2z}\mathcal T$ symmetry. Their intervalley interorbital pairing functions can be expressed as

\[
\Delta_{12}^{+-}(\mathbf k)
=\Delta_{\mathbf k}e^{i\varphi_{\mathbf k}},
\qquad
\Delta_{21}^{+-}(\mathbf k)
=\Delta_{\mathbf k}e^{-i\varphi_{\mathbf k}}.
\]

\begin{figure*}[h]
\centering
    \includegraphics[width=0.9\textwidth]{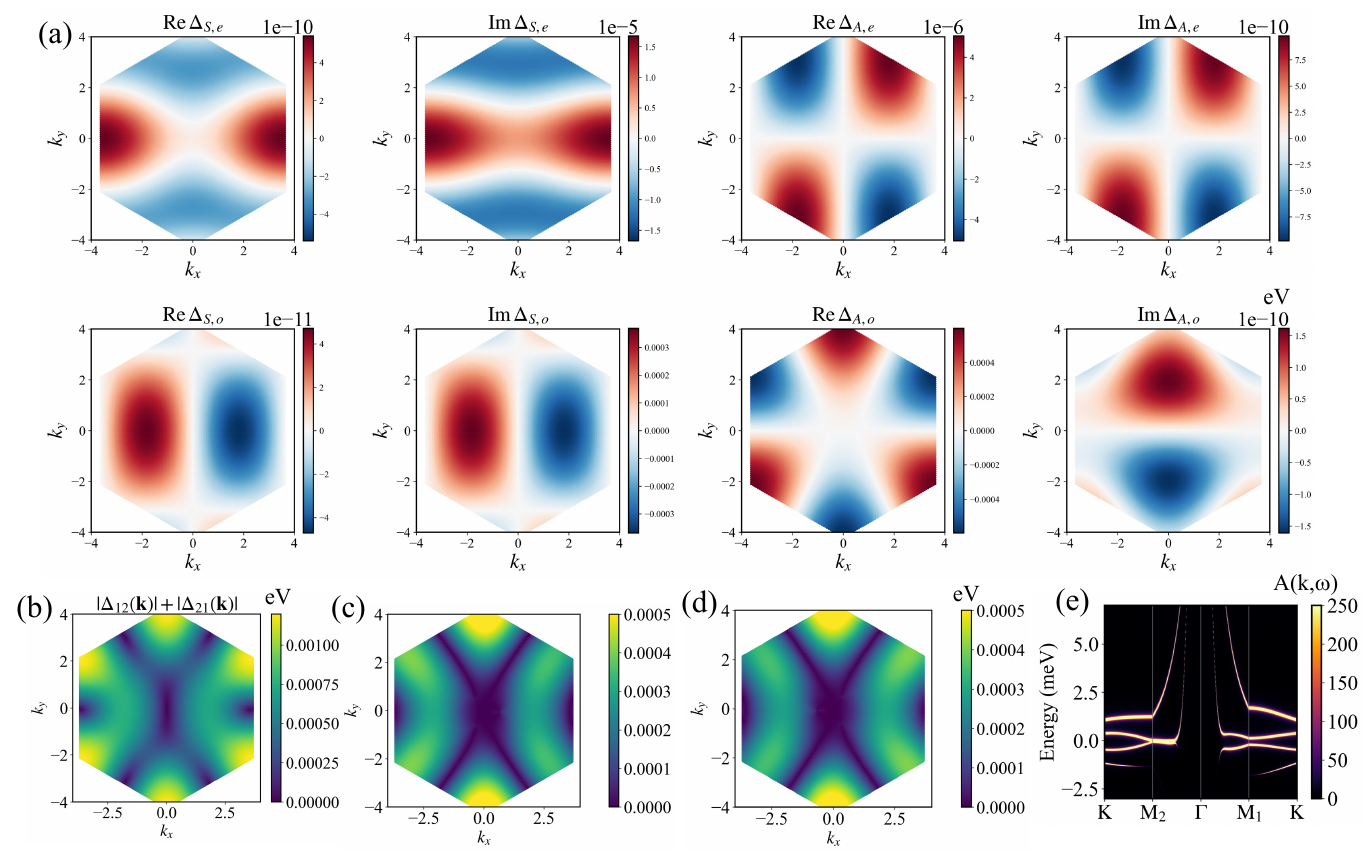}
\caption{Same as Fig.~\ref{fig:SC1pairing}, but for SC3 pairing state.}
\label{fig:SC3pairing}
\end{figure*}

\begin{figure*}[h]
\centering
    \includegraphics[width=0.9\textwidth]{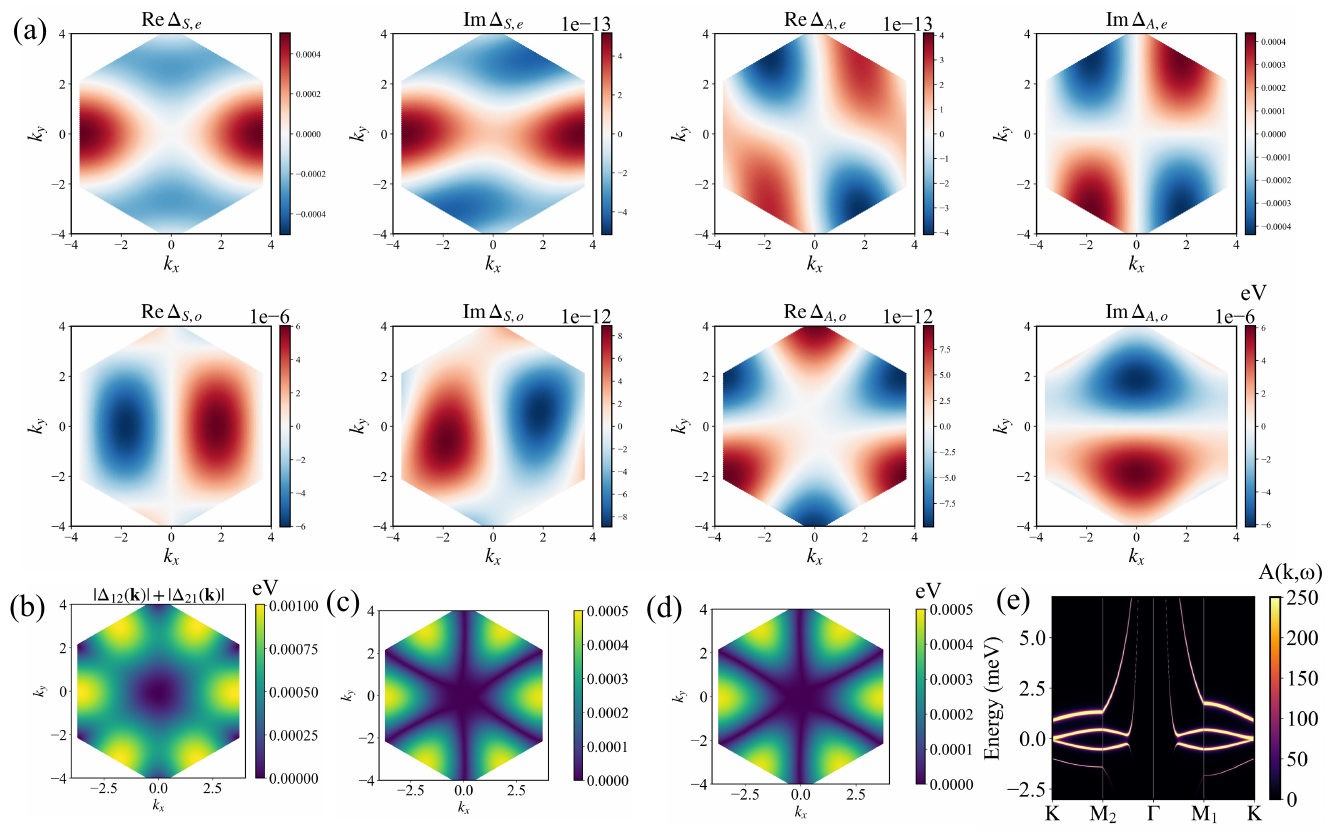}
\caption{Same as Fig.~\ref{fig:SC1pairing}, but for SC4 pairing state.}
\label{fig:SC4pairing}
\end{figure*}

\begin{figure*}[h]
\centering
    \includegraphics[width=0.9\textwidth]{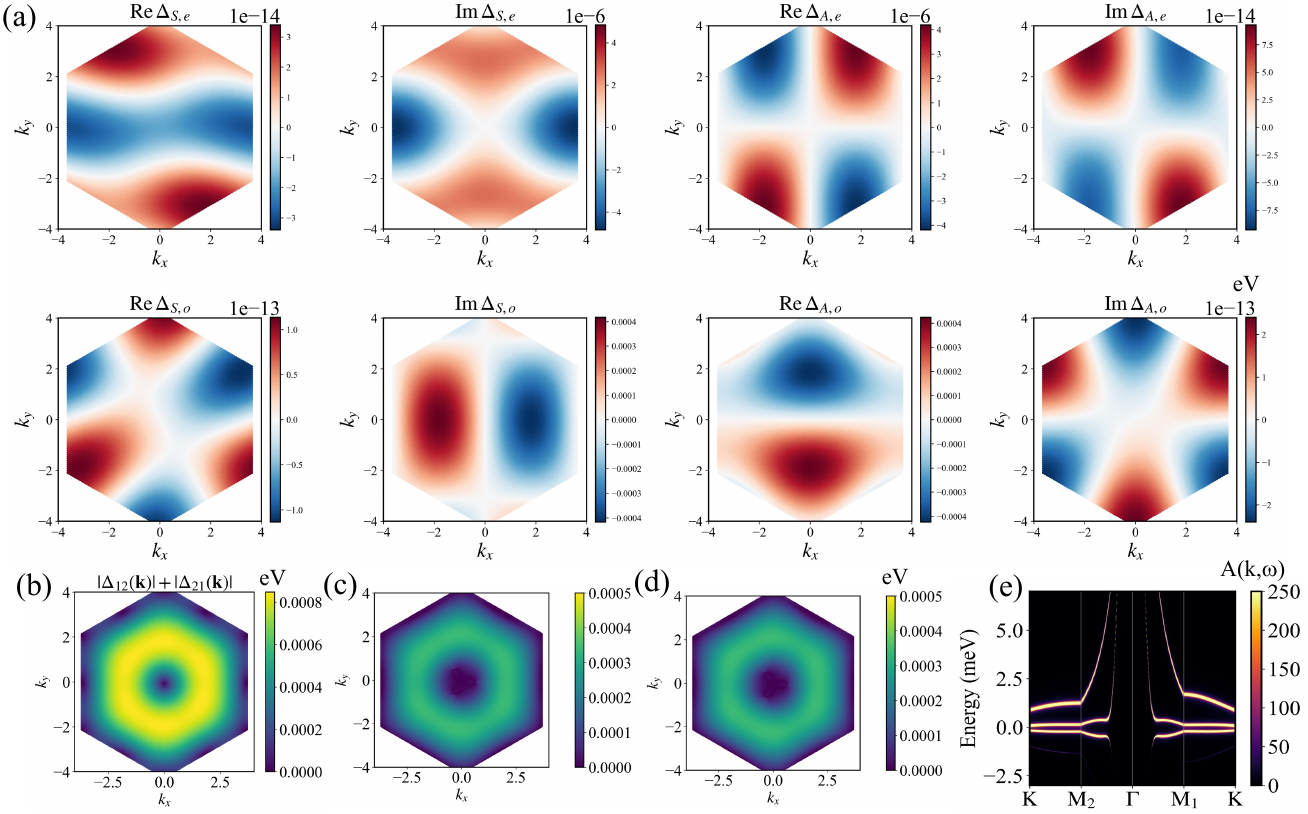}
\caption{Same as Fig.~\ref{fig:SC1pairing}, but for SC5 pairing state.}
\label{fig:SC5pairing}
\end{figure*}

\begin{figure*}[h]
\centering
\includegraphics[width=0.9\textwidth]{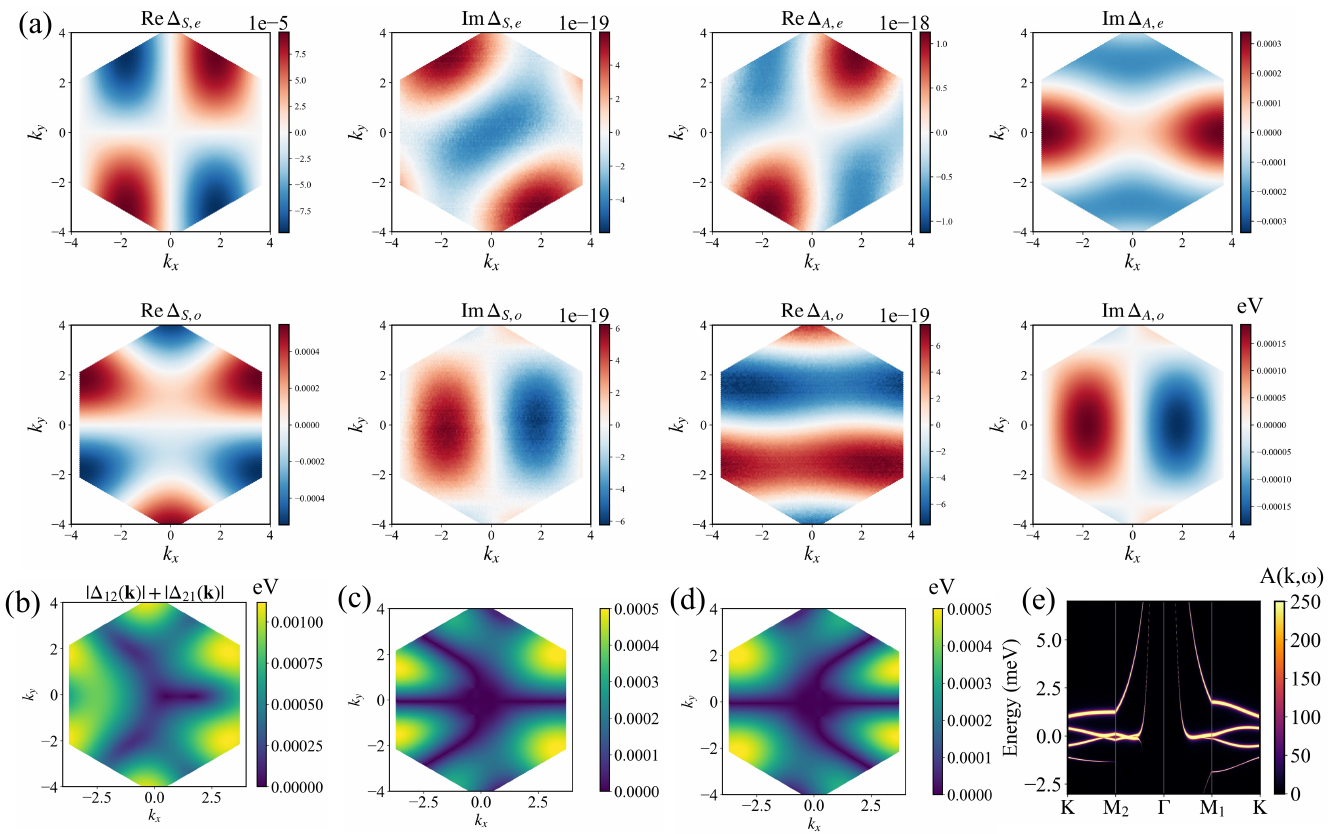}
\caption{Same as Fig.~\ref{fig:SC1pairing}, but for SC6 pairing state.}
\label{fig:SC6pairing}
\end{figure*}

The SC3 state has $\Delta_{12}$ and $\Delta_{21}$ of $p-if$ and $p+if$ form, respectively (Fig.~\ref{fig:SC3pairing}), and breaks $C_{3z}$ while preserving $C_{2z}$ and $C_{2x}$, both with eigenvalue $+1$ . Projection of the self-consistent gap yields approximately $10\%$ $A_1$ and $90\%$ $E_2$ character near $\nu = -2.9$.
%and $\nu = -1.9$. 
SC4 and SC5 states preserve $C_{3z}$ and consist of opposite chiral $d$-wave and $p$-wave components respectively in the two interorbital pairing channels, i.e.,

\[
\Delta_{12}^{+-}(\mathbf k)\sim d+id,
\qquad
\Delta_{21}^{+-}(\mathbf k)\sim d-id,
\]

for the SC4 (Fig.~\ref{fig:SC4pairing}) and 

\[
\Delta_{12}^{+-}(\mathbf k)\sim p-ip,
\qquad
\Delta_{21}^{+-}(\mathbf k)\sim p+ip.
\]

in the SC5 (Fig.~\ref{fig:SC5pairing}) state. Under $C_{3z}$, the orbital gap functions in both cases transform as

\[
\begin{aligned}
\Delta_{12}^{+-}(\mathbf k)
&\rightarrow
e^{i4\pi/3}
\Delta_{12}^{+-}(C_{3z}^{-1}\mathbf k)
=\Delta_{12}^{+-}(\mathbf k),\\
\Delta_{21}^{+-}(\mathbf k)
&\rightarrow
e^{-i4\pi/3}
\Delta_{21}^{+-}(C_{3z}^{-1}\mathbf k)
=\Delta_{21}^{+-}(\mathbf k),
\end{aligned}
\]
and hence belong to the $A_1$ representation. Finally, SC6 breaks $C_{3z}$, but transforms with eigenvalue $-1$ under $C_{2x}$. Together with its $C_{2z}=+1$ character, this restricts the allowed representations to $A_2$ and the $C_{2x}$-odd component of $E_2$. Its pairing structure contains a mixture of chiral $d$-wave and odd-momentum $f$- and $p$-wave harmonics (Fig.~\ref{fig:SC6pairing}),

\[
\begin{aligned}
\Delta_{12}^{+-}(\mathbf k)&\sim d+id+f-ip_x,\\
\Delta_{21}^{+-}(\mathbf k)&\sim d-id+f+ip_x.
\end{aligned}
\]

Projection onto the complete-gap irreducible representations gives approximately \(5\%\) \(A_2\) and \(95\%\) \(E_2\) character at $\nu=-2.9$ and \(10\%\) \(A_2\) and \(90\%\) \(E_2\) near $\nu = -1.9$. We note that the admixture of different irreps decreases with increasing temperature, and near $T_c$ we obtain two types of self-consistent solutions, belonging to the $A_1$ and $E_2$ representations.

\subsection{Intraorbital nearest-neighbor pairing}
\begin{figure*}[h]
\centering
    \includegraphics[width=0.5\textwidth]{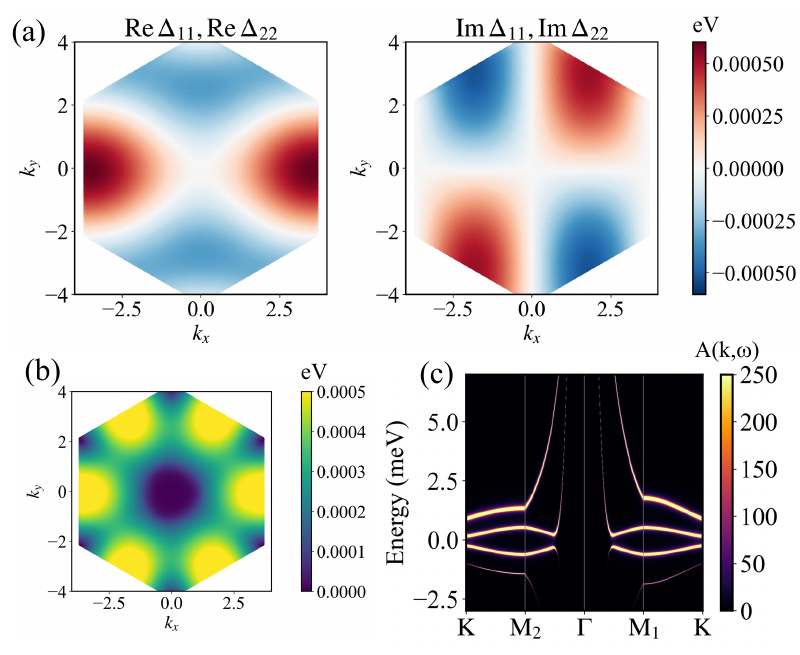}
\caption{Intraorbital NN pairing: orbital and momentum-parity decomposition of SC7 pairing. (a) Real and imaginary components of $\Delta_{11}$ and $\Delta_{22}$. (b) Magnitude of the band projected gap on the Fermi surface. (c) $f$-electron spectral function along the high-symmetry path $K$ - $M_2$ - $\Gamma$ - $M_1$ - $K$.
}
\label{fig:SC7pairing}
\end{figure*}

For intra-$f$-orbital NN pairing, we obtain the self-consistent solutions SC7-SC9,
as shown in Table.~I in the main text. The lowest-energy chiral $d_{x^2-y^2}+id_{xy}$ state SC7
(Fig.~\ref{fig:SC7pairing}) is of the form $(\sigma_0\otimes\tau_x)\otimes E_2$ and
belongs to $E_2$, breaking $\mathcal{T}$ and $C_{2x}$. The extended $s$-wave SC8,
$(\sigma_0\otimes\tau_x)\otimes A_1$, belongs to $A_1$. SC9 has
$\Delta_{11}=\Delta_{22}$, with $f$-wave symmetry
preserving $\mathcal{T}$; it is of the form $(\sigma_0\otimes i\tau_y)\otimes B_2$ and
belongs to $A_2$.

\subsection{Gap minima in band-projected gaps}

As discussed in main text, several superconducting states exhibit zeros in corresponding band-projected gaps. We further note that the $K$-point zeros in the band-projected gaps of SC4 and SC5, and those of SC7, are consistent with the symmetry-constrained point nodes expected for intra-Chern intervalley pairing in the $A_1$ channel and inter-Chern intervalley pairing in the $E_2$ channel, respectively, previously discussed in the context of phonon-mediated superconductivity~\cite{Liu2024}.

\subsection{Onsite pairing}

\begin{figure*}[h]
\centering
    \includegraphics[width=0.5\textwidth]{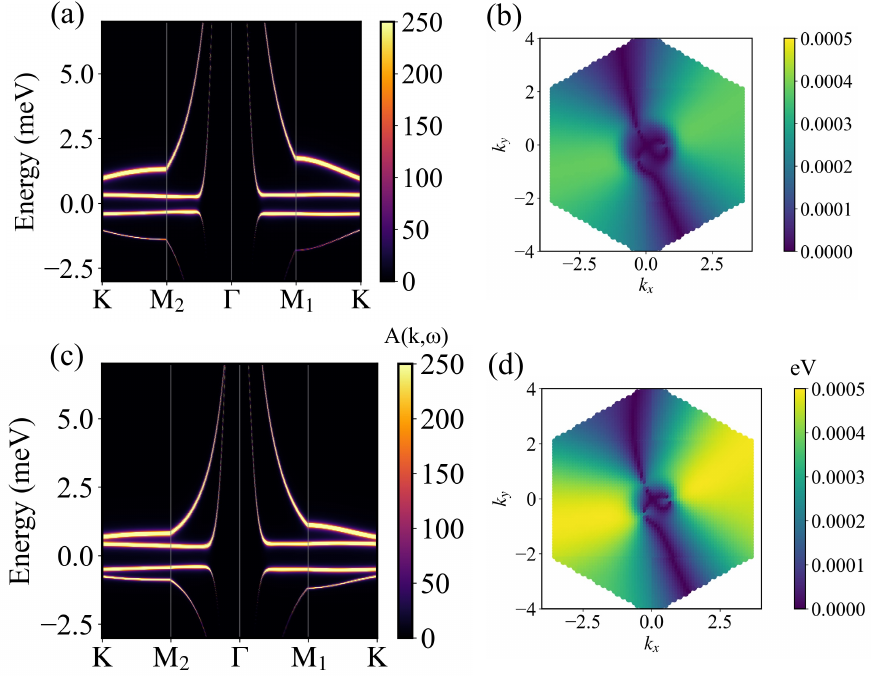}
\caption{Electron spectral functions $A(\mathbf{k}, \omega)$ for the onsite interorbital pairing with nematic $d$-wave symmetry at $\nu= -2.9$ (a) and $\nu = -1.9$ (c). Magnitude of band-projected gaps over the moiré Brillouin zone showing nematic $p$-wave like nodal structure at $\nu= -2.9$ (b) and $\nu = -1.9$ (d). }
\label{fig:figSMonsite}
\end{figure*}

For comparison with the NN pairing solutions, we also consider onsite pairing interactions in the $f$-electron sector. Intervalley spin-singlet interorbital pairing transforms as the $E_2$ representation, whereas intraorbital pairing belongs to the $A_1$ representation with $s$-wave symmetry. The two-component $E_2$ order parameter admits both chiral and nematic $d$-wave states~\cite{Wang2024}. Within our parameter regime, the self-consistent mean-field solution near the optimal fillings favors the nematic state, with $\Delta_{12}=\Delta e^{-i\phi}$ and $\Delta_{21}=\Delta e^{i\phi}$. The band-projected gap exhibits a nematic $p$-wave-like nodal structure, as illustrated by the Fermi-surface-projected gap and electronic spectral function $A(\mathbf{k},\omega)$ of the superconducting state in Fig.~\ref{fig:figSMonsite}.

\begin{figure*}[h]
\centering
    \includegraphics[width=0.9\textwidth]{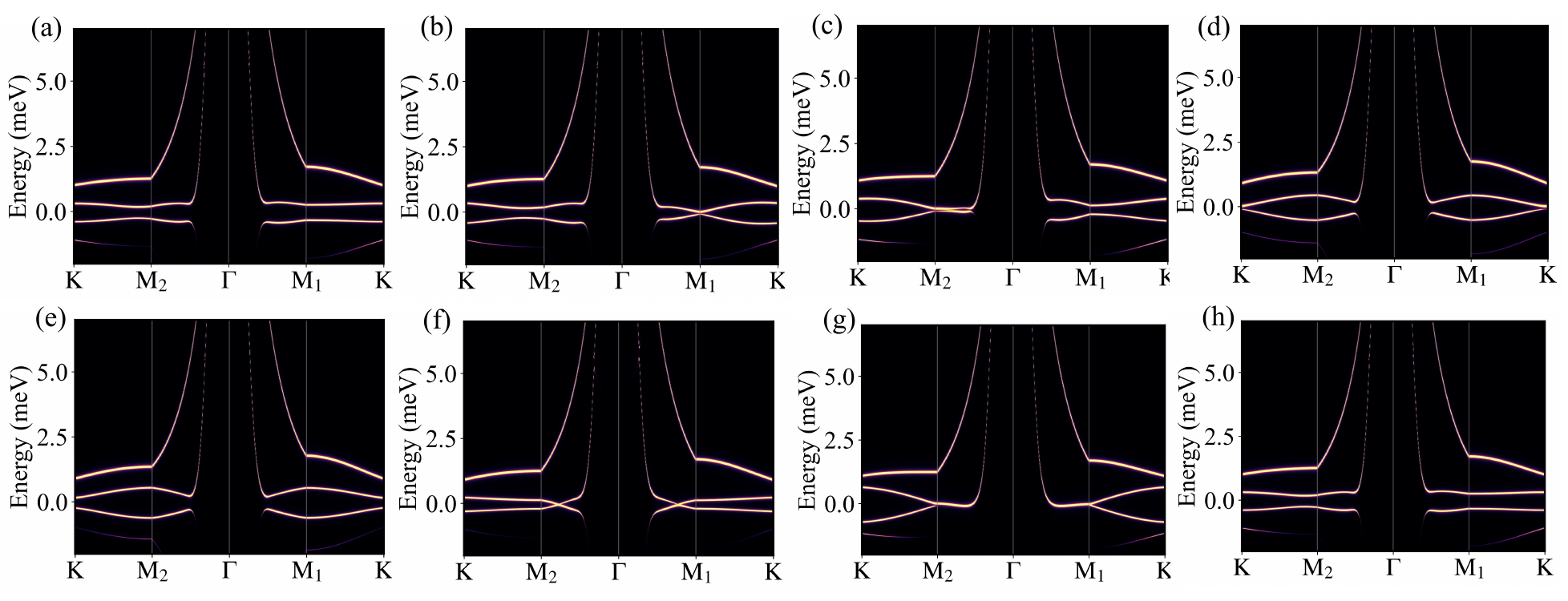}
\caption{Electron spectral functions $A(\mathbf{k},\omega)$ for (a) SC1, (b) SC2, (c) SC3, (d) SC4, (e) SC7, (f) SC8, (g) SC9, (h) SC6 states, along the high-symmetry path along $K$ - $M_2$ - $\Gamma$ - $M_1$ - $K$ at $\nu = -1.9$.}
\label{fig:figSM3}
\end{figure*}

\begin{figure*}[h]
\centering
    \includegraphics[width=0.95\textwidth]{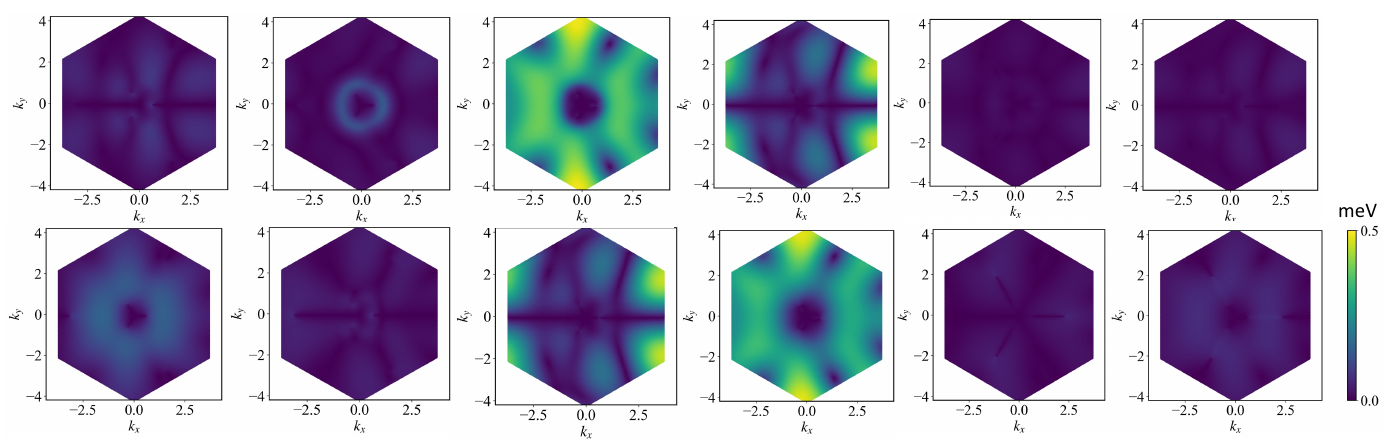}
\caption{Momentum-resolved band-projected superconducting pairing amplitudes \(|\Delta_{nm}^{\mathrm{band}}(\mathbf{k})|\) corresponding to SC1 nematic $d$+$p$ state at optimal filling. The upper and lower rows correspond to \(n=3\) and \(n=4\), respectively, and the columns correspond to \(m=1,\ldots,6\), from left to right. The two flat bands are \(n=3,4\).}
\label{fig:figSM-interband-sc1}
\end{figure*}

\begin{figure*}[h]
\centering
    \includegraphics[width=0.95\textwidth]{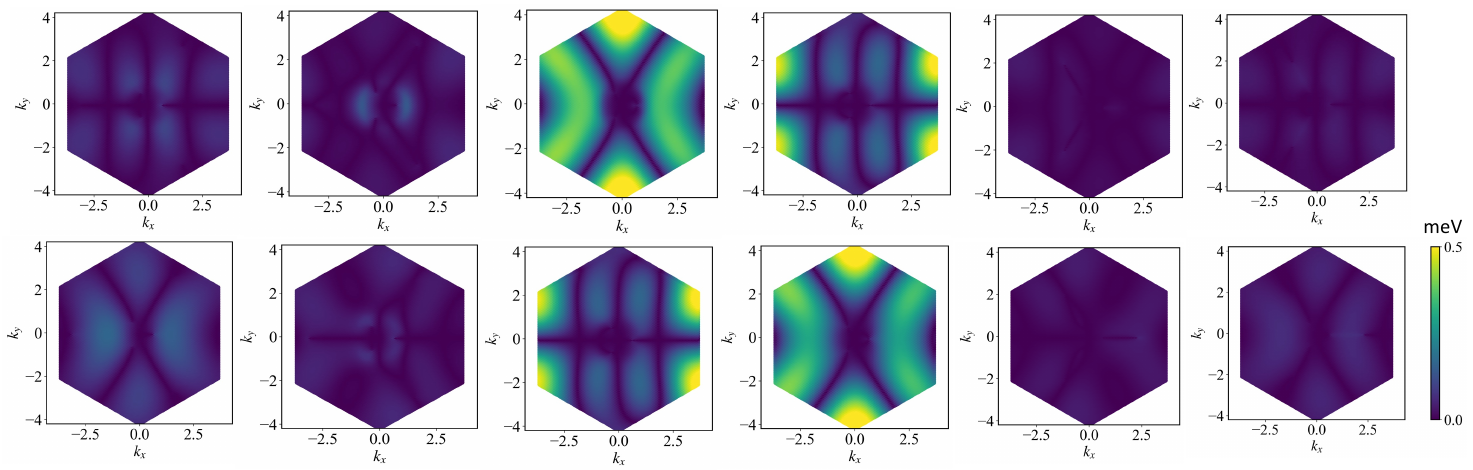}
\caption{Same as Fig.~\ref{fig:figSM-interband-sc1}, but with the SC state corresponding to SC3 nematic $p\pm if$ state.}
\label{fig:figSM-interband-sc3}
\end{figure*}

\begin{figure*}[h]
\centering              
\includegraphics[width=0.95\textwidth]{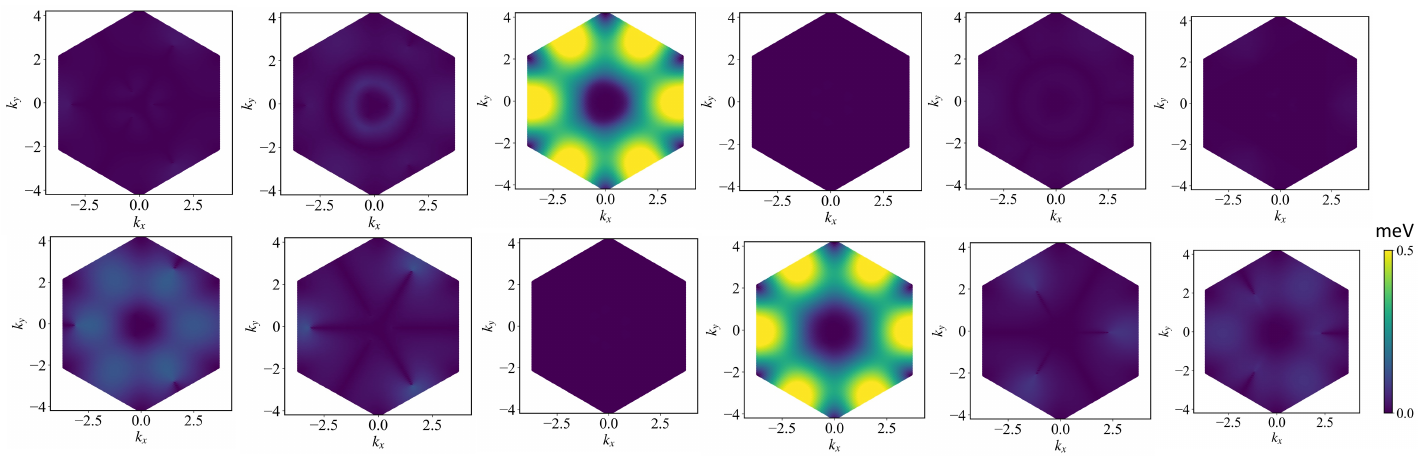}
\caption{Same as Fig.~\ref{fig:figSM-interband-sc1}, but with the SC state corresponding to SC7 chiral $d_{x^2-y^2}+ id_{xy}$ state.}
\label{fig:figSM-interband-sc5}
\end{figure*}

\section{Superfluid weight and intra/interband component decomposition}

The general formula for the superfluid weight from linear response theory is~\cite{lamponenSuperconductivityPairDensity2025, huhtinenRevisitingFlatBand2022} 

\begin{align}
\label{eq: supp_sfw_green_fn_form_first_line}
D^s_{\mu\nu}
&=
\frac{1}{V}
\sum_{\mathbf{k}}
\sum_{i\omega_n}
\mathrm{Tr}
\left[
\partial_{k_\mu}\partial_{k_\nu}\widetilde{H}\big|_{\mathbf{k}}
\,G(i\omega_n,\mathbf{k})
\right]
-
\frac{1}{U V_c}
\sum_{\alpha}
\left(
\frac{\delta \Delta_\alpha}{\delta A_\mu}
\frac{\delta \Delta_\alpha^*}{\delta A_\nu}
+
\mathrm{H.c.}\right) +
\\
\label{eq: supp_sfw_green_fn_form_second_line}
&\frac{1}{V}
\sum_{\mathbf{k}}
\sum_{i\omega_m}
\mathrm{Tr}
\Big[
G(i\omega_m,\mathbf{k})
\left(
\partial_{k_\mu}\widetilde{H}\big|_{\mathbf{k}+\mathbf{q}/2}\gamma^z
+
\delta_\mu \Delta
\right)
\nonumber\\
&\times
G(i\omega_m+i\nu_m,\mathbf{k}+\mathbf{q})
\left(
\partial_{k_\nu}\widetilde{H}\big|_{\mathbf{k}+\mathbf{q}/2}\gamma^z
+
\delta_\nu \Delta
\right)
\Big],
\end{align}
where $E_{\bk a}$ and $|\psi_{\bk a}\rangle$ are the eigenvalues and eigenvectors of the BdG Hamiltonian, respectively, $\gamma^z$ is a Pauli matrix in the particle-hole space, $\Vec{\Delta}$ is a vector of the mean-field order parameters in orbital basis, and $V_c = V/N$ is the volume of the unit cell, with $V$ the volume and $N$ the number of unit cells of the system. Here,

\begin{align}
\tilde{H}(\mathbf{k})
\equiv
\begin{pmatrix}
h^+_{\rm eff}(\mathbf{k}) & 0 \\
0 & -h^{-*}_{\rm eff}(-\mathbf{k})
\end{pmatrix}.
\end{align}
and the variation of the order parameter with respect to the vector potential $\bm{A}$ is encoded in  

\begin{align}
    \delta_\mu\Delta_{\bm{k}}&\equiv\begin{pmatrix}0&\frac{\delta\Delta_{\bm{k}}}{\delta      
    A_\mu}\\\frac{\delta \Delta^{\dagger}_{\bm{k}}}{\delta A_\mu}&0\end{pmatrix}.
\end{align}

The first line (Eq.~\eqref{eq: supp_sfw_green_fn_form_first_line}) contains the diamagnetic term (first term) and the term arising from the $\bm{A}$-dependence of the self-consistent gap. The second and third lines are the paramagnetic current–current correlator, along with additional terms due to the $\bm{A}$-dependence of the gap. We find that the gap-response terms are numerically indistinguishable from zero, across all fillings and temperatures, and hence we will drop them from now on. For a detailed discussion of these terms, see \cite{huhtinenRevisitingFlatBand2022, lamponenSuperconductivityPairDensity2025}.

\subsection{Intra- and interband components}

Using Eqs.~\eqref{eq: supp_sfw_green_fn_form_first_line} and \eqref{eq: supp_sfw_green_fn_form_second_line}, dropping the terms associated to the $\bm A$-dependence of the gap, and using integration by parts in the diamagnetic contribution, we have the following form of $D^{s}_{\mu\nu}$, with $G (i\omega_m, \bk) = \sum_n\frac{|\psi_{\bk n}\rangle\langle\psi_{\bk n}|}{i\omega_m - E_{\bk n}}$:
 
\begin{align}
    D^{s}_{\mu\nu}
        =&\frac{1}{V}\sum_{\bm{k},a,b}\frac{n_{F}(E_{\bk a})-n_{F}(E_{\bk b})}{E_{\bk b}-E_{\bk a}}\Bigg[ \langle\psi_{\bk a}|(\partial_{{\mu}}\widetilde{H}|_{\bm{k}})|\psi_{\bk b}\rangle\langle\psi_{\bk b}|(\partial_{{\nu}} {H}_{\mathrm{BdG}}(\bk ))|\psi_{\bk a}\rangle \label{eq: rvb sfw 1} \\
        &- \langle\psi_{\bk a}|\partial_{{\mu}}\widetilde{H}|_{\bm{k}}\gamma^{z}|\psi_{\bk b}\rangle\langle\psi_{\bk b}|\partial_{{\nu}}\widetilde{H}|_{\bm{k}}\gamma^{z}|\psi_{\bk a}\rangle \Bigg] . \label{eq: rvb sfw 2}
\end{align}

In obtaining Eq.~\eqref{eq: rvb sfw 2}, we have subtracted a boundary contribution $B_{\mu\nu}=\oint_{\partial\mathrm{BZ}}\frac{dS}{(2\pi)^2}\,n_\mu\sum_n f(E_n)\ \partial_{k_\nu}E_n$, with $E_n$ being the corresponding eigenvalues, $f$ is the Fermi function, $n_\mu$ is the $\mu$ component of the unit vector in outward normal direction. $B_{\mu\nu}$ originates because the $c$-electron sector of the THF model is not periodic across the moiré Brillouin-zone boundary~\cite{Song2022}. Such a boundary contribution is also present in the normal state and hence to obtain a physical zero $ D^{s}_{\mu\nu}$ in the normal state, this boundary contribution is substracted in the literature ~\cite{kopnin2010,Liang2017}. We have also verified numerically that $|\delta B_{xx}=B_{xx}^{\mathrm{SC}}-B_{xx}^{\mathrm N}|$ is at most
approximately $6\%$ of $D^s_{xx}$ in the superconducting state at optimal filling $\nu = -2.9$. Moreover, adding $B_{\mu\nu}$ in Eq.~\eqref{eq: rvb sfw 2} do not change the qualitative findings, including the doping and temperature dependence. Hence, we use Eq.~\eqref{eq: rvb sfw 2} in order calculate $D^{s}_{\mu\nu}$ throughout the manuscript. Eq.~\eqref{eq: rvb sfw 2} has an added advantage as it helps to separate $D^s_{xx}$ into intra- and interband contributions.

To split the superfluid weight into its intra- and interband components (with respect to the normal state bands), we expand the eigenvectors $|\phi_{\bk a}\rangle$ in terms of the Bloch functions $|n_{\boldsymbol{k}\sigma}\rangle$ (eigenvectors of the Hamiltonian without the attractive interaction)\cite{Liang2017, lamponenSuperconductivityPairDensity2025} 

\begin{equation}\label{nsbandproj}
|\psi_{\mathbf{k}a}\rangle
=
\sum_{n=1}^{N_{\rm orb}}
\left[
w_{+,an\mathbf{k}}\,|+\rangle\otimes|n_{\mathbf{k}+}\rangle
+
w_{-,an\mathbf{k}}\,|-\rangle\otimes|n^{*}_{-\mathbf{k}-}\rangle
\right].
\end{equation}
where $|\pm \rangle$ are the eigenvectors of $\gamma^z$ with eigenvalues $\pm 1$, and $N_{\mathrm{orb}}$ is the total number of orbitals (both $f$ and $c$). With this, the superfluid weight becomes

\begin{align}
    D_{\mu \nu } &=  \frac{2}{V}\sum_{\bk} \sum_{nmls} C_{++--}^{nmls}[j_{\mu}^{+}(\boldsymbol{k})]_{nm}[j_{\nu}^{-}(-\boldsymbol{k})]_{sl} +(\mu\leftrightarrow\nu) \label{eq: sfw_pure_cur_op}\\
    &+ \frac{2}{V} \sum_{\bk}\sum_{nmls} \text{Re} \bigg[ \langle n_{\bk +}| \partial_{\mu}\Delta_{\bk} | m^*_{-\bk -}\rangle (C^{nmls}_{+-++} [j_{\nu}^{+}(\boldsymbol{k})]_{ls} + C^{nmls}_{+---} [j_{\nu}^{-}(\boldsymbol{-k})]_{sl} \bigg] \label{eq: sfw_delta_k_term}
\end{align}
where
\begin{align}
C_{p_1 p_2 p_3 p_4}^{nmls}&=\sum_{ab}\frac{n_{F}(E_{\bk a})-n_{F}(E_{\bk b})}{E_{\bk b}-E_{\bk a}}w_{p_1,an\bk}^{*}w_{p_2,bm\boldsymbol{k}}w_{p_3,bl\boldsymbol{k}}^{*}w_{p_4,as\boldsymbol{k}},\\
[j_{\mu}^{\eta}(\boldsymbol{k})]_{nm}&=\langle n_{\boldsymbol{k}\eta}|\partial_{k_{\mu}}H_{\boldsymbol{k}}^{\eta}|m_{\boldsymbol{k}\eta}\rangle\\
&=\delta_{nm}\partial_{{\mu}}\epsilon_{\boldsymbol{k}n\eta}+(\epsilon_{\boldsymbol{k}n\eta}-\epsilon_{\boldsymbol{k}m\eta})\langle\partial_{{\mu}}n_{\boldsymbol{k}\eta}|m_{\boldsymbol{k}\eta}\rangle,
\end{align}
with $p_i \in \{ +, -\}$. Here $\Delta_\mathbf{k}$ is the order parameter matrix in orbital space, and $\partial_\mu \equiv \partial_{k_\mu}$. Equation~\eqref{eq: sfw_pure_cur_op} is written purely in terms of the current operator $\bm{j}$, while the terms in Eq.~\eqref{eq: sfw_delta_k_term} are due to the $\bk$-dependence of the order parameters. It should be clarified that this $\bk$-dependence, resulting from the nontrivial Fourier transformation of the real-space order parameters in the case with non-onsite interactions, is distinct from the previously discarded $\bm{A}$-dependence of the order parameters, which quantifies the change of the self-consistent values of the real-space order parameters as a vector potential is applied to the system. In this work, we define the intraband component of the superfluid weight to be the part of these sums with $m = n$ and $l = s$, while taking everything else to be the interband component. Note that the interband component thus contains both fully interband terms with $m \neq n$ and $l \neq s$, as well as mixed terms such as $m = n$, $l \neq s$.

This decomposition of the superfluid weight is closely related to the often-used split into conventional and geometric parts. Sometimes, the notions of a "gap term" and/or "multigap term" appear as well, describing the terms due to the $\bk$-dependence of the order parameters; see \cite{Liang2017, lamponenSuperconductivityPairDensity2025, kitamuraSuperconductivityMonolayerFeSe2022} for more discussion. Here, we opt to use the terms intra- and interband with the aforementioned definitions for simplicity.

We note that the gap term, Eq.~\eqref{eq: sfw_delta_k_term}, is different from the corresponding equation (H12) of \cite{lamponenSuperconductivityPairDensity2025}, as it was recently noticed by one of the authors of both the present work and of \cite{lamponenSuperconductivityPairDensity2025} that Eq. (H12) of \cite{lamponenSuperconductivityPairDensity2025} is incorrect. We have numerically verified Eq.\eqref{eq: sfw_delta_k_term} to be the correct form.

\subsection{Quantum-geometric origin of the
intraband current}

Here, we discuss the relation between the
Hartree-induced intraband current and the quantum
geometry of the flat bands. To establish this
connection analytically, we consider the projection
of the Hartree potential onto the bare flat-band
subspace of the THF model. We write the single-particle Hamiltonian as
\begin{equation}
H(\mathbf{k})=H_0(\mathbf{k})+V_H,
\end{equation}
where $H_0(\mathbf{k})$ is the non-interacting part of the THF Hamiltonian, defined in Eq.~\eqref{eq:H0}, and
$V_H$ contains the orbital-dependent
Hartree shifts. In the orbital basis
$(f_1,f_2,c_1,c_2,c_3,c_4)$, the static
potential can be written as
\begin{equation}
V_H=
\epsilon_f P_f+
\epsilon_{c12}P_{c12}+
\epsilon_{c34}P_{c34},
\end{equation}
where $P_f$, $P_{c12}$, and $P_{c34}$
are projectors onto the corresponding
orbital sectors. The quantities
$\epsilon_f$, $\epsilon_{c12}$, and
$\epsilon_{c34}$ denote the respective
diagonal energy shifts. For the quasiparticle Hamiltonian of Eq.~\eqref{nsham},
$H_{\rm eff}=\sqrt{\hat Z}\,(H_0+V_H)\sqrt{\hat Z}$, the same analysis applies with
$H_0\to\sqrt{\hat Z}H_0\sqrt{\hat Z}$ and $V_H\to\sqrt{\hat Z}V_H\sqrt{\hat Z}$.

We further consider the ideal flat-band limit
of the THF model, neglecting the bare
flat-band dispersion and the splitting
generated by the parameter $M$ (see Eq.~(\ref{cHamiltonianMatrix})).
The two bare flat bands
have vanishing weight on the $c_1,c_2$
orbitals and consist only of the
$f$ and $c_3,c_4$ components.
Considering
\begin{equation}
P_0(\mathbf{k})
=
\sum_{a=1}^{2}
|u_a^{(0)}(\mathbf{k})\rangle
\langle u_a^{(0)}(\mathbf{k})|
\end{equation}
to denote the projector onto the two bare
flat bands of $H_0(\mathbf{k})$, and
$Q_0(\mathbf{k})=I-P_0(\mathbf{k})$
the projector onto the 
remote-band subspace, and neglecting the $c_1,c_2$ orbital weight
in the bare flat bands, we have
\begin{equation}
P_0P_{c12}P_0\simeq 0,
\end{equation}
and consequently,
\begin{equation}
P_0P_{c34}P_0
\simeq P_0-P_0P_fP_0.
\end{equation}
Thus, the Hartree potential projected
onto the bare flat-band subspace becomes
\begin{align}
V_H^{\mathrm{proj}}(\mathbf{k})
&=P_0(\mathbf{k})V_HP_0(\mathbf{k})
\nonumber\\
&\simeq
\epsilon_{c34}P_0(\mathbf{k})
+
(\epsilon_f-\epsilon_{c34})
P_0(\mathbf{k})P_fP_0(\mathbf{k}).
\label{eq:projected_hartree}
\end{align}
Hereafter, we denote
$\epsilon_c\equiv\epsilon_{c34}$. Within this approximation, 
the momentum dependence of the projected
Hamiltonian arises only from the projected
Hartree potential. Its eigenvalues are
therefore approximately
\begin{equation}
\epsilon_{n\mathbf{k}+}
\simeq
\epsilon_c+
(\epsilon_f-\epsilon_c)
w_{nf}(\mathbf{k}),
\label{eq:projected_energy}
\end{equation}
where
\begin{equation}
w_{nf}(\mathbf{k})
=
\langle\widetilde u_n(\mathbf{k})|
P_f|\widetilde u_n(\mathbf{k})\rangle
\end{equation}
is the $f$-orbital weight of the
eigenstate $|\widetilde u_n\rangle$
obtained by diagonalizing the projected
Hartree Hamiltonian.

Consequently, the Hartree-induced
intraband velocity becomes
\begin{equation}
v_{n,\mu}(\mathbf{k})
\simeq
(\epsilon_f-\epsilon_c)
\partial_{k_\mu}w_{nf}(\mathbf{k}).
\label{eq:hartree_velocity}
\end{equation}
The corresponding contribution to the
product of diagonal band velocities is
\begin{equation}
\mathcal V_{n,\mu\mu}(\mathbf{k})
\simeq
(\epsilon_f-\epsilon_c)^2
\left(\partial_{k_\mu}w_{nf}(\mathbf{k})\right)^2.
\label{eq:hartree_velocity_product}
\end{equation}

We now show that the orbital-weight
gradient is constrained by the quantum
geometry of the bare flat-band subspace.
Within the projected model,
$|\widetilde u_n\rangle$ satisfies
\begin{equation}
P_0P_fP_0|\widetilde u_n(\mathbf{k})\rangle
=
w_{nf}(\mathbf{k})|\widetilde u_n(\mathbf{k})\rangle.
\end{equation}
Taking the momentum derivative gives
\begin{align}
\partial_{k_\mu}w_{nf}(\mathbf{k})
&=
\langle\widetilde u_n(\mathbf{k})|
\partial_{k_\mu}(P_0P_fP_0)
|\widetilde u_n(\mathbf{k})\rangle
\nonumber\\
&=
2\operatorname{Re}
\langle \partial_{k_\mu}
\widetilde u_n(\mathbf{k})|
Q_0P_f|\widetilde u_n(\mathbf{k})\rangle.
\label{eq:weight_gradient}
\end{align}
Here, we used
$P_0|\widetilde u_n(\mathbf{k})\rangle
=|\widetilde u_n(\mathbf{k})\rangle$
and
$P_0(\partial_{k_\mu}P_0)P_0=0$. Applying the Cauchy-Schwarz inequality
to Eq.~\eqref{eq:weight_gradient} (and using $Q_0^2=Q_0$),
we obtain
\begin{align}
\left(\partial_{k_\mu}w_{nf}(\mathbf{k})\right)^2
&\leq
4\langle\partial_{k_\mu}\widetilde u_n(\mathbf{k})|
Q_0|\partial_{k_\mu}\widetilde u_n(\mathbf{k})\rangle
\nonumber\\
&\quad\times
\langle\widetilde u_n(\mathbf{k})|
P_fQ_0P_f|\widetilde u_n(\mathbf{k})\rangle.
\end{align}
Since the states $|\widetilde{u}_m(\mathbf{k})\rangle$ diagonalize
$P_0 P_f P_0$, we have
$\langle\widetilde{u}_m(\mathbf{k})|P_f|\widetilde{u}_n(\mathbf{k})\rangle
= w_{nf}(\mathbf{k})\delta_{mn}$, and therefore
\begin{align}
\langle\widetilde u_n(\mathbf{k})|P_f Q_0 P_f|\widetilde u_n\rangle
&=\langle\widetilde u_n(\mathbf{k})|P_f^2|\widetilde u_n\rangle
\nonumber\\
&\quad-\langle\widetilde u_n(\mathbf{k})|P_f P_0 P_f|\widetilde u_n(\mathbf{k})\rangle
\nonumber\\
&=w_{nf}(\mathbf{k})-\sum_m
\left|\langle\widetilde u_m|P_f|\widetilde u_n\rangle\right|^2
\nonumber\\
&=w_{nf}(\mathbf{k})-w_{nf}^2(\mathbf{k})
=w_{nf}(\mathbf{k})(1-w_{nf}(\mathbf{k})).
\end{align}

Defining the quantum metric contribution
associated with overlap of variation of the bare
flat-band subspace with the remote band subspace
\begin{equation}
g_{n,\mu\mu}(\mathbf{k})
=
\langle\partial_{k_\mu}\widetilde u_n(\mathbf{k})|
Q_0|
\partial_{k_\mu}\widetilde u_n(\mathbf{k})\rangle.
\label{eq:subspace_metric}
\end{equation}
we arrive at
\begin{equation}
\left(\partial_{k_\mu}w_{nf}(\mathbf{k})\right)^2
\leq
4w_{nf}(\mathbf{k})(1-w_{nf}(\mathbf{k}))
g_{n,\mu\mu}(\mathbf{k}).
\label{eq:metric_bound}
\end{equation}

\begin{figure*}[h]
\centering              \includegraphics[width=0.85\textwidth]{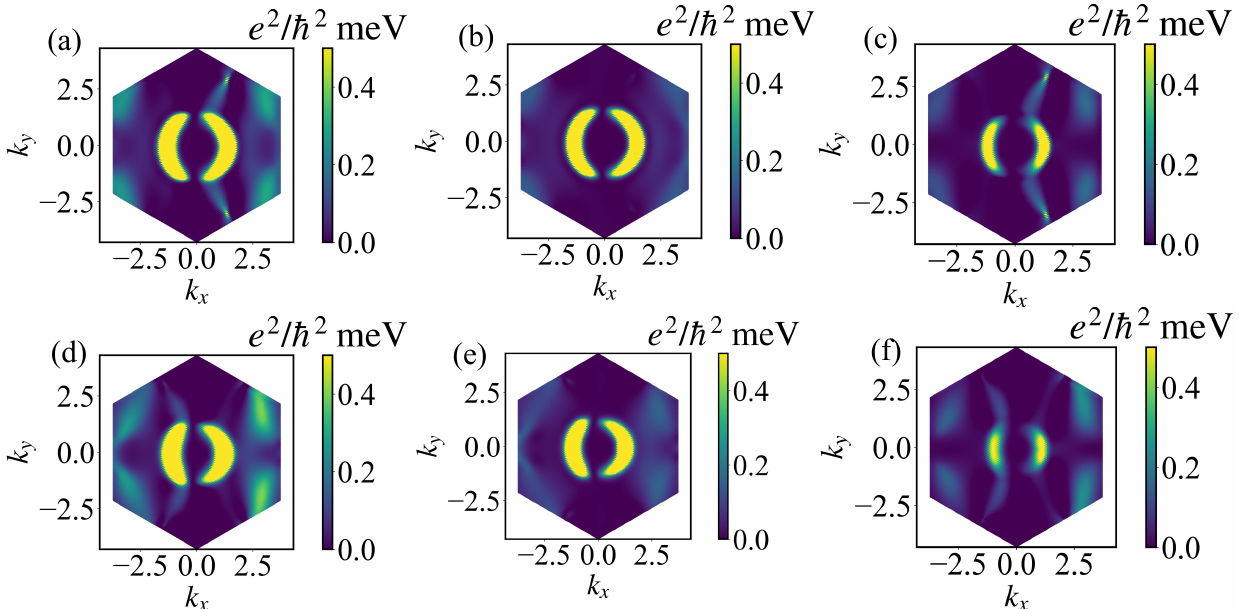}
\caption{Extended data: Momentum-resolved superfluid-weight integrands at $\nu=-2.9$ (upper row) and $\nu=-1.9$ (lower row). Integrands of the $D^s_{xx}$ (a,d),$D^{s,intra}_{xx}$ (b,e), and $D^{s,inter}_{xx}$ (c,f) in the moiré Brillouin zone. Both $D^{s,intra}_{xx}$ and $D^{s,inter}$ are concentrated in similar momentum-space regions near $\Gamma$.}
\label{fig:figSM-Dskspace}
\end{figure*}

\begin{figure*}
\centering
    \includegraphics[width=0.5\textwidth]{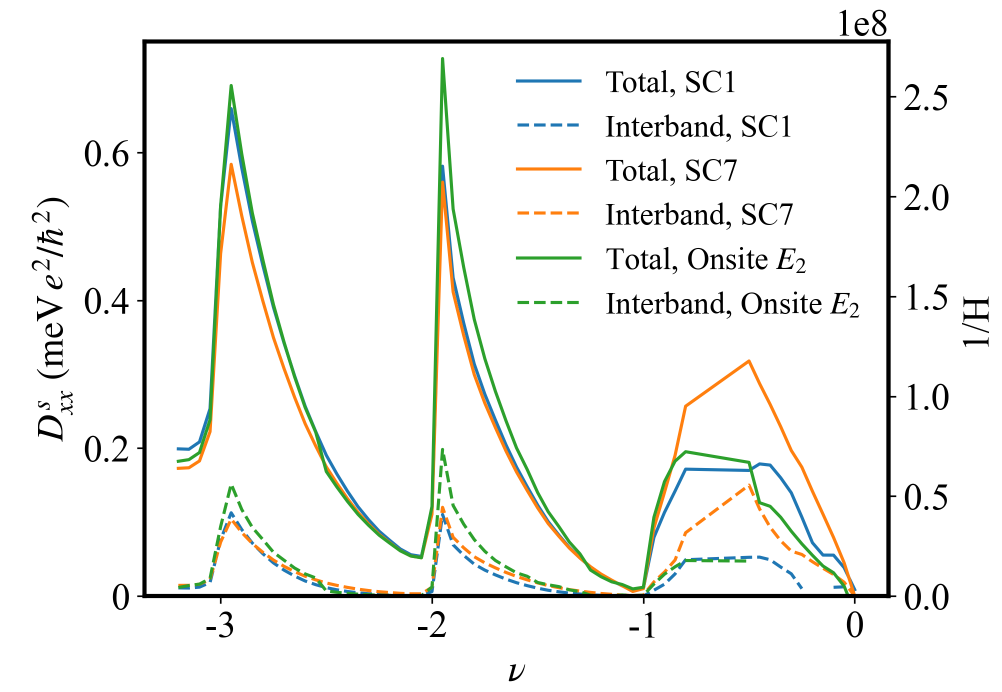}
\caption{Doping dependence of the superfluid weight $D^s_{xx}$ and its interband contribution for SC1, SC7, and onsite $E_2$ pairing. }
\label{fig:smdopingDs}
\end{figure*}

\begin{figure*}
\centering
    \includegraphics[width=0.4\textwidth]{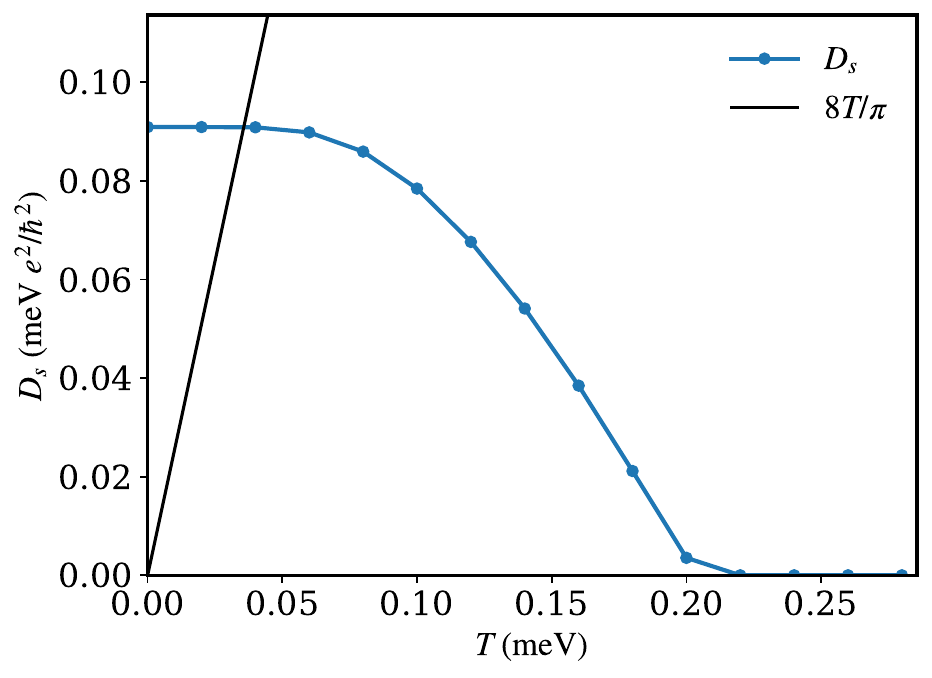}
\caption{Temperature dependence of the superfluid weight with with a model containing only $f$ orbitals resulting $T_{\rm BKT}\simeq 0.4K$}
\label{fig:figSM-2}
\end{figure*}

\begin{figure*}
\centering
    \includegraphics[width=0.45\textwidth]{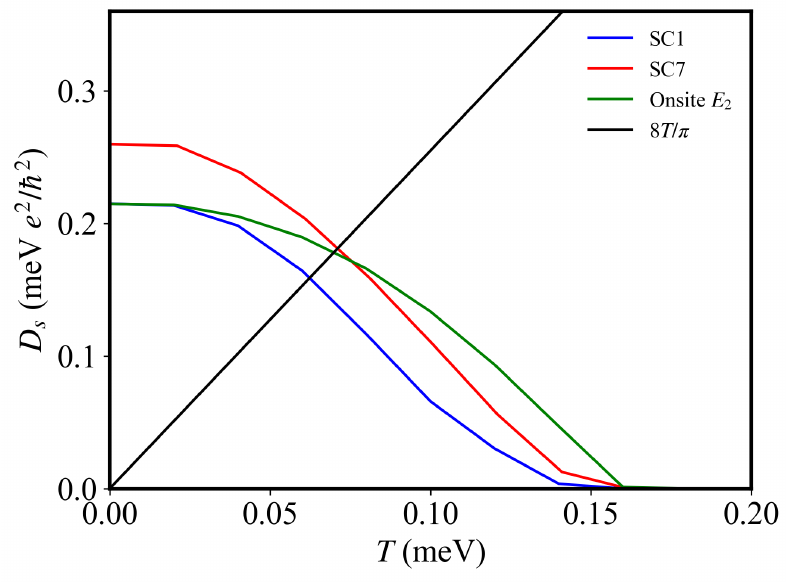}
\caption{Superfluid weight $D_s$ as a function of temperature near $\nu\approx -0.8$ for different pairing symmetries, with the estimated BKT transition temperatures around $0.7$-$0.8$K.}
\label{fig:sm3}
\end{figure*}

\begin{figure*}
\centering
    \includegraphics[width=0.4\textwidth]{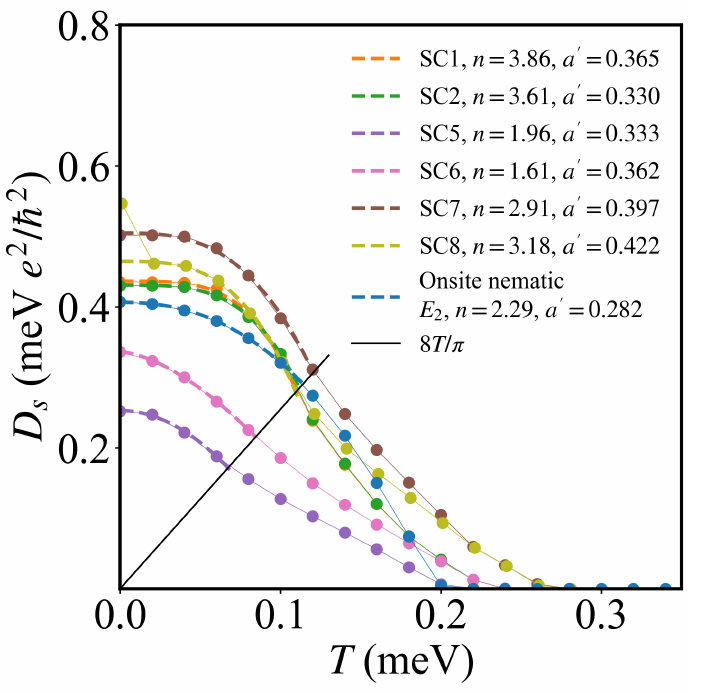}
\caption{Temperature dependence of $D_s$ for different SC states and power-law fits below $T_{\mathrm{BKT}}$ using $D_s(T)=D_s(0)-aT^n$ at $\nu=-1.9$. Here $n$ is the fitted exponent and $a'=aT_{\mathrm{BKT}}^n/D_s(0)$.}
\label{fig:smdopingDs}
\end{figure*}

\begin{figure*}
\centering
    \includegraphics[width=0.85\textwidth]{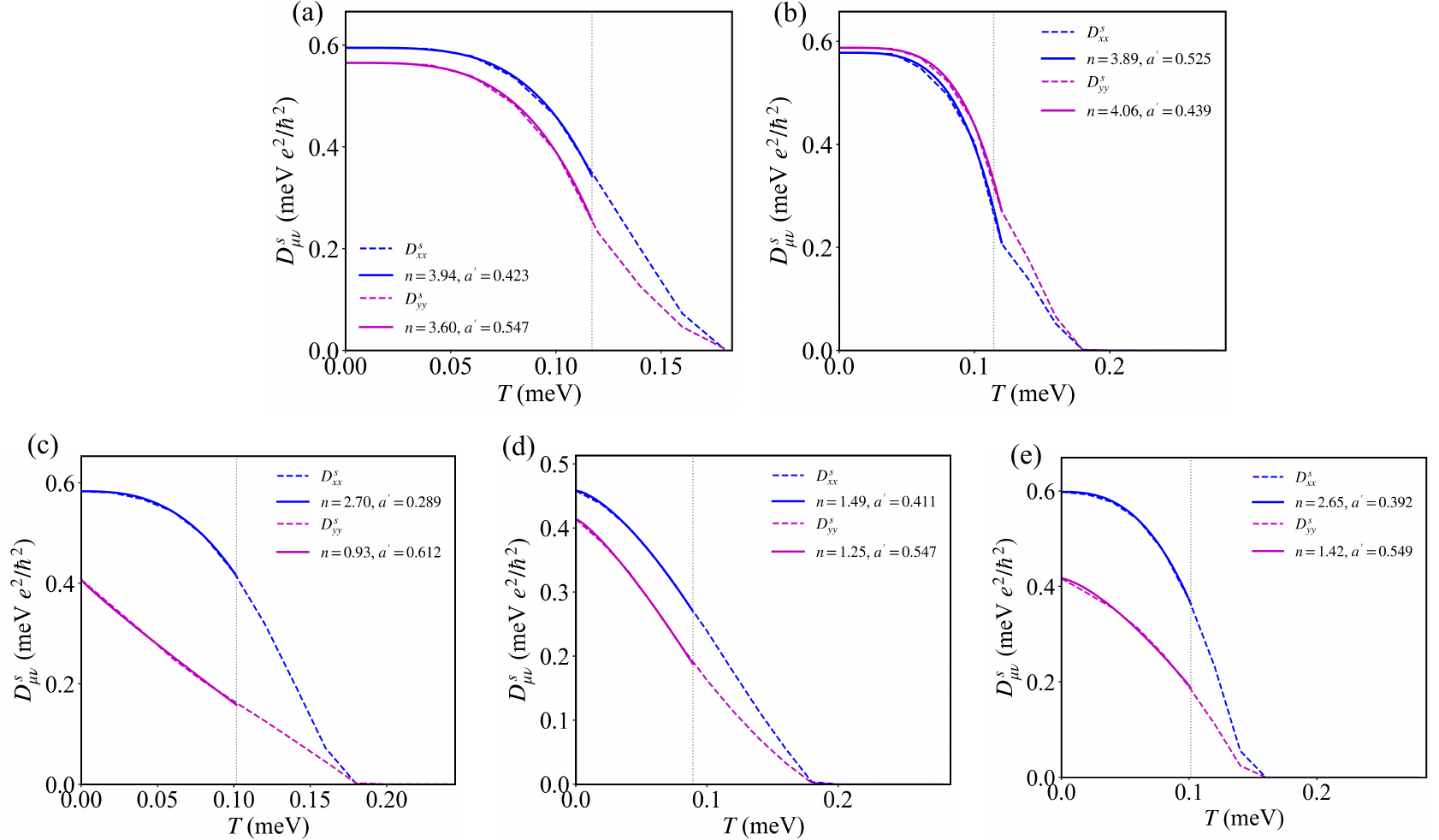}
\caption{$D^{s}_{xx}$ and $D^{s}_{yy}$ for $C_{3z}$ breaking SC states SC1 (a), SC2 (b), SC3 (c), SC6 (d) and onsite pairing (e) at $\nu = -2.9$.}
\label{fig:sm2}
\end{figure*}

\begin{table*}[t]
\centering
\small
\renewcommand{\arraystretch}{1.15}
\begin{tabular}{llcccc}
\hline\hline
State & Fitting window & $n\pm\sigma_n$
& $a'\pm\sigma_{a'}$ & $R^2$ & RMSE \\
\hline
SC1 & $T<T_{\mathrm{BKT}}$
& $3.860\pm0.338$ & $0.3653\pm0.0178$ & 0.996349 & 0.002360 \\
SC2 & $T<T_{\mathrm{BKT}}$
& $3.606\pm0.127$ & $0.3305\pm0.0064$ & 0.999398 & 0.000865 \\
SC5 & $T<T_{\mathrm{BKT}}$
& $1.960\pm0.287$ & $0.3334\pm0.0207$ & 0.996630 & 0.001470 \\
SC6 & $T<T_{\mathrm{BKT}}$
& $1.606\pm0.037$ & $0.3616\pm0.0031$ & 0.999824 & 0.000534 \\
SC7 & $T<T_{\mathrm{BKT}}$
& $2.909\pm0.174$ & $0.3974\pm0.0091$ & 0.997533 & 0.003383 \\
SC8 & $T<T_{\mathrm{BKT}}$
& $3.185\pm0.209$ & $0.4224\pm0.0124$ & 0.998750 & 0.001863 \\
\hline\hline
\end{tabular}
\par\smallskip
\caption{
Power-law fits at $\nu=-1.9$ to $D_s(T)=D_s(0)-aT^n$,
with $a$ and $n$ being free parameters, and
$a'=aT_{\mathrm{BKT}}^n/D_s(0)$.
$\sigma_n$ and $\sigma_{a'}$ denote the standard errors
of the fitted parameters $n$ and $a'$, respectively.
RMSE denotes the root mean square error.
The same data have also been fitted with the exponential
BCS form
$D_s(T)=D_s(0)[1-\sqrt{2\pi\Delta_0/T}
\exp(-\Delta_0/T)]$,
with $\Delta_0$ as the fitting parameter.
In all cases, the power-law fit yields a smaller RMSE
and a larger $R^2$ than the exponential BCS fit.
}
\label{powerlawfitsn1p9}
\end{table*}

\begin{table*}[t]
\centering
\small
\renewcommand{\arraystretch}{1.15}
\begin{tabular}{llcccc}
\hline\hline
State & Fitting window & $n\pm\sigma_n$
& $a'\pm\sigma_{a'}$ & $R^2$ & RMSE \\
\hline
SC1 & $T<T_{\mathrm{BKT}}$
& $3.688\pm0.113$ & $0.4849\pm0.0106$ & 0.999545 & 0.001208 \\

SC2 & $T<T_{\mathrm{BKT}}$
& $3.569\pm0.119$ & $0.4472\pm0.0093$ & 0.999463 & 0.001356 \\

SC3 & $T<T_{\mathrm{BKT}}$
& $1.293\pm0.016$ & $0.4611\pm0.0013$ & 0.999978 & 0.000315 \\

SC4 & $T<T_{\mathrm{BKT}}$
& $1.517\pm0.036$ & $0.4859\pm0.0056$ & 0.999900 & 0.000472 \\

SC5 & $T<T_{\mathrm{BKT}}$
& $3.519\pm0.055$ & $0.4541\pm0.0031$ & 0.999832 & 0.001019 \\

SC6 & $T<T_{\mathrm{BKT}}$
& $1.357\pm0.032$ & $0.4824\pm0.0045$ & 0.999799 & 0.000922 \\

SC7 & $T<T_{\mathrm{BKT}}$
& $2.627\pm0.224$ & $0.4795\pm0.0236$ & 0.996378 & 0.003566 \\

SC8 & $T<T_{\mathrm{BKT}}$
& $2.027\pm0.201$ & $0.5414\pm0.0273$ & 0.995155 & 0.006167 \\
\hline\hline
\end{tabular}
\par\smallskip
\caption{Same as Table~\ref{powerlawfitsn1p9}, but for
$\nu=-2.9$. For SC7, the exponential BCS fit yields
a smaller RMSE and a larger $R^2$ than the power-law fit.}
\label{tab:powerlaw_fits_n2p9}
\end{table*}

\bibliographystyle{apsrev4-1}
%\bibliography{tbg,tbg_gap,sm,Intro}

\end{document}